\documentclass[a4paper,draft]{amsart} 
\usepackage{amsmath}
\usepackage{epsfig}
\usepackage{amssymb} 

\begin{document}

\title{ Hyperelliptic Theta-Functions and Spectral Methods } 
\author{J.~Frauendiener}
\address{Institut f\"ur Astronomie und Astrophysik, Universit\"at T\"ubingen,
Auf der Morgenstelle 10, 72076 T\"ubingen, Germany}
\email{joergf@tat.physik.uni-tuebingen.de}

\author{C.~Klein}
\address{LUTh, Observatoire de Paris, 92195 Meudon Cedex, France}
\curraddr{Max-Planck-Institut f\"ur Physik, F\"ohringer Ring 6,
    80805 M\"unchen, Germany}
    \email{klein@mppmu.mpg.de}
\date{\today}    

\begin{abstract}
   A code for the numerical evaluation of hyperelliptic
   theta-func\-tions is presented. Characteristic quantities of the
   underlying Riemann surface such as its periods are determined with the
   help of spectral methods. The code is optimized for solutions of
   the Ernst equation where the branch points of the Riemann surface
   are parameterized by the physical coordinates. An exploration of the
   whole parameter space of the solution is thus only possible with an
   efficient code. The use of spectral approximations allows for
   an efficient calculation of all quantities in the solution with high
   precision. The case of almost degenerate Riemann surfaces is
   addressed. Tests of the numerics using identities for periods on
   the Riemann surface and integral identities for the Ernst potential
   and its derivatives are performed. It is shown that an accuracy of
   the order of machine precision can be achieved. These accurate
   solutions are used to provide boundary conditions for a code which
   solves the axisymmetric stationary Einstein equations. The resulting
   solution agrees with the theta-functional solution to very high precision.
\end{abstract}

\maketitle

\section{Introduction}

Solutions to integrable differential equations in terms of
theta-functions were introduced with the works of Novikov, Dubrovin,
Matveev, Its, Krichever, \ldots(see~\cite{DubNov75,ItsMat75,Kric78,algebro}) for
the Korteweg-de Vries (KdV) equation. Such solutions to e.g.\ the KdV,
the Sine-Gordon, and the Non-linear Schr\"odinger equation describe
periodic or quasi-periodic solutions,
see~\cite{dubrovin81,algebro}. They are given explicitly in terms of
Riemann theta-functions defined on some Riemann surface. Though all
quantities entering the solution are in general given in explicit form
via integrals on the Riemann surface, the work with theta-functional
solutions admittedly has not reached the importance of soliton
solutions.

The main reason for the more widespread use of solitons is that they
are given in terms of algebraic or exponential functions. On the other
hand the parameterization of theta-functions by the underlying Riemann
surface is very implicit. The main parameters, typically the branch
points of the Riemann surface, enter the solutions as parameters in
integrals on the Riemann surface. A full understanding of the
functional dependence on these parameters seems to be only possible
numerically. In recent years algorithms have been developed to
establish such relations for rather general Riemann surfaces as
in~\cite{tretkoff84} or via Schottky uniformization (see 
\cite{algebro}), which have been incorporated successively in
numerical and symbolic codes,
see~\cite{seppala94,hoeij94,gianni98,deconinck01,deconinck03} and
references therein (the last two references are distributed along with
Maple~6, respectively Maple~8, and as a Java implementation at 
\cite{riemann}). For an approach to express periods of hyperelliptic 
Riemann surfaces via theta constants see \cite{enoric2003}.

These codes are convenient to study theta-functional solutions of
equations of KdV-type where the considered Riemann surfaces are
`static', i.e., independent of the physical coordinates. In these
cases the characteristic quantities of the Riemann surface have to be
calculated once, just the comparatively fast summation in the
approximation of the theta series via a finite sum as e.g.\
in~\cite{deconinck03} has to be carried out in dependence of the
space-time coordinates.

The purpose of this article is to study numerically theta-functional
solutions of the Ernst equation~\cite{ernst} which were given by
Korotkin~\cite{Koro88}. In this case the branch points of the
underlying hyperelliptic Riemann surface are parameterized by the
physical coordinates, the spectral curve of the Ernst equation is in
this sense `dynamical'. The solutions are thus not studied on a single
Riemann surface but on a whole family of surfaces. This implies that
the time-consuming calculation of the periods of the Riemann surface
has to be carried out for each point in the space-time. This includes
limiting cases where the surface is almost degenerate. In addition the
theta-functional solutions should be calculated to high precision in
order to be able to test numerical solutions for rapidly rotating
neutron stars such as provided e.g. by the spectral code
\texttt{LORENE}~\cite{Lorene}. This requires a very efficient code of high
precision.

We present here a numerical code for hyperelliptic surfaces where the
integrals entering the solution are calculated by expanding the
integrands with a Fast Cosine Transformation in MATLAB. The precision
of the numerical evaluation is tested by checking identities for
periods on Riemann surfaces and by comparison with exact
solutions. The code is in principle able to deal with general
(non-singular) hyperelliptic surfaces, but is optimized for a genus~2
solution to the Ernst equation which was constructed
in~\cite{prl2,prd3}. We show that an accuracy of the order of machine
precision ($\sim 10^{-14}$) can be achieved at a space-time point in
general position with 32 polynomials and in the case of almost
degenerate surfaces which occurs e.g., when the point approaches the
symmetry axis with at most 256 polynomials.  Global tests of the
numerical accuracy of the solutions to the Ernst equation are provided
by integral identities for the Ernst potential and its derivatives:
the equality of the Arnowitt-Deser-Misner (ADM) mass and the Komar
mass (see~\cite{komar,wald}) and a generalization of the Newtonian virial
theorem as derived in~\cite{virial}. We use the so determined numerical data
for the theta-functions to provide `exact' boundary values on a sphere
for the program library \texttt{LORENE}~\cite{Lorene} which was developed for a
numerical treatment of rapidly rotating neutron stars. \texttt{LORENE} solves
the boundary value problem for the stationary axisymmetric Einstein
equations with spectral methods. We show that the theta-functional
solution is reproduced to the order of $10^{-11}$ and better.

The paper is organized as follows: in section~\ref{sec:ernsteq} we
collect useful facts on the Ernst equation and hyperelliptic Riemann
surfaces, in section~\ref{sec:spectral} we summarize basic features of
spectral methods and explain our implementation of various
quantities. The calculation of the periods of the hyperelliptic
surface and the non-Abelian line integrals entering the solution is
performed together with tests of the precision of the numerics. In
section~\ref{sec:integrals} we check integral identities for the Ernst
potential. The test of the spectral code \texttt{LORENE} is presented
in section~\ref{sec:lorene}. In section~\ref{sec:concl} we add some
concluding remarks.

\section{Ernst equation and hyperelliptic Riemann surfaces}
\label{sec:ernsteq}

The Ernst equation for the complex valued potential $\mathcal{E}$ (we 
denote the real and the imaginary part of $\mathcal{E}$ with $f$ and 
$b$ respectively)
depending on the two coordinates $(\rho,\zeta)$ can be written in the
form
\begin{equation}
    \Re \mathcal{E}\left(\mathcal{E}_{\rho\rho}+\frac{1}{\rho}
    \mathcal{E}_{\rho}+\mathcal{E}_{\zeta\zeta}\right)=
    \mathcal{E}_{\rho}^{2}+\mathcal{E}_{\zeta}^{2}
    \label{ernst1}.
\end{equation}
The equation has a physical interpretation as the stationary
axisymmetric Einstein equations in vacuum (see appendix and references
given therein). Its complete integrability was shown by
Maison~\cite{maison} and Belinski-Zakharov~\cite{belzak}. For real
Ernst potential, the Ernst equation reduces to the axisymmetric
Laplace equation for $\ln \mathcal{E}$. The corresponding solutions
are static and belong to the so called Weyl class, see~\cite{exac}.

Algebro-geometric solutions to the Ernst equation were given by
Korotkin~\cite{Koro88}. The solutions are defined on a family of
hyperelliptic surfaces $\mathcal{L}(\xi,\bar{\xi})$ with
$\xi=\zeta-i\rho$ corresponding to the plane algebraic curve
\begin{equation}
    \mu^{2}=(K-\xi)(K-\bar{\xi})\prod_{i=1}^{g}(K-E_{i})(K-F_{i})
    \label{hyper1},
\end{equation}
where $g$ is the genus of the surface and where the branch points
$E_{i}$, $F_{i}$ are independent of the physical coordinates and for
each $n$ subject to the reality condition $E_{n}=\bar{F}_{n}$ or
$E_{n},F_{n}\in \mathbb{R}$.

Hyperelliptic Riemann surfaces are important since they show up in the
context of algebro-geometric solutions of various integrable equations
as KdV, Sine-Gordon and Ernst. Whereas it is a non-trivial problem to
find a basis for the holomorphic differentials on general surfaces
(see e.g.~\cite{deconinck01}), it is given in the hyperelliptic case
(see e.g.~\cite{algebro}) by
\begin{equation}
d\nu_k = \left(   \frac{dK}{\mu}, \frac{KdK}{\mu},\ldots,
  \frac{K^{g-1}dK}{\mu} \right)
    \label{basis},
\end{equation}
which is the main simplification in the use of these surfaces.  We
introduce on $\mathcal{L}$ a canonical basis of cycles
$(a_{k},b_{k})$, $k=1,\ldots,n$. The holomorphic differentials
$d\omega_k$ are normalized by the condition on the $a$-periods
\begin{equation}
    \int_{a_{l}}^{}d\omega_{k}=2\pi i \delta_{lk}.
    \label{normholo}
\end{equation}
The matrix of $b$-periods is given by $\mathbf{B}_{ik} =
\int_{b_{i}}^{}d\omega_{k}$. The matrix $\mathbf{B}$ is a so-called
Riemann matrix, i.e.\ it is symmetric and has a negative definite real
part. The Abel map $\omega: \mathcal{L} \to \mbox{Jac}(\mathcal{L}) $ with base
point $E_{1}$ is defined as $\omega(P)=\int_{E_{1}}^{P}d\omega_k$, 
where $\mbox{Jac}(\mathcal{L})$is the Jacobian of $\mathcal{L}$.  The
theta-function with characteristics corresponding to the curve
$\mathcal{L}$ is given by
\begin{equation}
    \Theta_{\mathbf{p}\mathbf{q}}(\mathbf{x}|\mathbf{B})=
    \sum_{\mathbf{n}\in\mathbb{Z}^{g}}^{}\exp\left\{\frac{1}{2}
    \langle\mathbf{B}(\mathbf{p}+\mathbf{n}),(\mathbf{p}+\mathbf{n})
    \rangle+\langle\mathbf{p}+\mathbf{n},2i\pi\mathbf{q}+\mathbf{x}
    \rangle\right\}
    \label{theta},
\end{equation}
where $\mathbf{x}\in \mathbb{C}^{g}$ is the argument and
$\mathbf{p},\mathbf{q}\in \mathbb{C}^{g}$ are the characteristics. We
will only consider half-integer characteristics in the following. The
theta-function with characteristics is, up to an exponential factor,
equivalent to the theta-function with zero characteristic (the Riemann
theta-function is denoted with $\Theta$) and shifted argument,
\begin{equation}
    \Theta_{\mathbf{p}\mathbf{q}}(\mathbf{x}|\mathbf{B})=
    \Theta(\mathbf{x}+\mathbf{B}\mathbf{p}+2i\pi\mathbf{q})\exp\left\{
    \frac{1}{2}\langle\mathbf{B}\mathbf{p},\mathbf{p}
    \rangle+\langle\mathbf{p},2i\pi\mathbf{q}+\mathbf{x} \rangle\right\}.
    \label{theta2}
\end{equation}

We denote by $d\omega_{PQ}$ a differential of the third kind, i.e., a
1-form which has poles in $P,Q\in \mathcal{L}$ with respective. residues
$+1$ and $-1$. This singularity structure characterizes the
differentials only up to an arbitrary linear combination of 
holomorphic differentials. The meromorphic differentials  can be
normalized by the condition that all $a$-periods vanish. We use the
notation $\infty^{\pm}$ for the infinite points on different sheets of
the curve $\mathcal{L}$, namely $\mu/K^{g+1}\to \pm 1$ as $K\to
\infty^{\pm}$. The differential $d\omega_{\infty^{+}\infty^{-}}$ is
given up to holomorphic differentials by $-K^{g}dK/\mu$. It is well
known that the $b$-periods of normalized differentials of the third
kind can be expressed in terms of the Abel map (see
e.g.~\cite{dubrovin81}),
\begin{equation}
    \int_{b_{k}}^{}d\omega_{PQ}=\omega_{k}(P)-\omega_{k}(Q), \quad
    k=1,\ldots,g
    \label{period}.
\end{equation}

In~\cite{prl,prd2} a physically interesting subclass of Korotkin's
solution was identified which can be written in the form
\begin{equation}
    \mathcal{E}=\frac{\Theta_{\mathbf{p}\mathbf{q}}(\omega(\infty^{+})+\mathbf{u})}{
    \Theta_{\mathbf{p}\mathbf{q}}(\omega(\infty^{-})+\mathbf{u})}\cdot
  e^{I}
    \label{ernst2},
\end{equation}
where $\mathbf{u}=(u_k)\in\mathbb{C}^g$ and where
\begin{equation}
    I=\frac{1}{2\pi i}\int_{\Gamma}^{}\ln G(K)\,d\omega_{\infty^{+}
    \infty^{-}}(K), \qquad u_k=\frac{1}{2\pi i}
    \int_{\Gamma}^{}\ln G(K)\,d\omega_k.
    \label{path1}
\end{equation}
$\Gamma$ is a piece-wise smooth contour on $\mathcal{L}$ and $G(K)$ is
a non-zero H\"older-continuous function on $\Gamma$. The contour
$\Gamma$ and the function $G$ have to satisfy the reality conditions
that with $K\in \Gamma$ also $\bar{K}\in \Gamma$ and
$\bar{G}(\bar{K})=G(K)$; both are independent of the physical
coordinates.

In the following we will discuss the example of the solution
constructed in~\cite{prl2,prd3} which can be interpreted as a disk of
collisionless matter. For a physical interpretation
see~\cite{prd4}. The solution is given on a surface of the
form~(\ref{hyper1}) with genus 2. The branch points independent of the
physical coordinates are related through the relations
$E_{i}=\bar{F}_{i}$, $i=1,2$ and $E_{1}=-F_{2}$. The branch points are
parameterized by two real parameters $\lambda$ and $\delta$. Writing
$E_{1}^{2}=\alpha +i\beta$ with real $\alpha$, $\beta$, we have
\begin{equation}
    \alpha=-1+\frac{\delta}{2}, \quad
    \beta=\sqrt{\frac{1}{\lambda^{2}} +\delta-\frac{\delta^{2}}{4}}
    \label{disk1}.
\end{equation}
The contour $\Gamma$ is the piece of the covering of the imaginary
axis in the upper sheet between $[-i,i]$, the function $G$ has the
form
\begin{equation}
    G(K)=\frac{\sqrt{(K^{2}-\alpha)^{2}+\beta^{2}}+K^{2}+1}{
    \sqrt{(K^{2}-\alpha)^{2}+\beta^{2}}-K^{2}-1}.
    \label{disk2}
\end{equation}
The physical parameters vary between $\delta=0$, the solution which
was first given in~\cite{NeuMei95}, and
$\delta_{s}=2(1+\sqrt{1+1/\lambda^{2}})$, the static limit in which
$\beta=0$. In the latter case the Riemann surface degenerates, the
resulting Ernst potential (\ref{ernst2}) is real and be expressed in terms 
of objects corresponding to the surface
$\mathcal{L}_{0}$ of genus 0 defined by the relation
$\mu_{0}^{2}=(K-\xi)(K-\bar{\xi})$. The parameter $\lambda$ varies
between $\lambda=0$, the so-called Newtonian limit where the branch
points $E_{i}$, $F_{i}$ tend to infinity. Since $G$ is also of order
$\lambda$ in this limit, the lowest order contributions are again real
and defined on the surface $\mathcal{L}_{0}$. This case corresponds to
the disk limit of the Maclaurin ellipsoids, see~\cite{bitr}. The upper
limit for $\lambda$ is infinity for $\delta\neq 0$ and
$\lambda_{c}=4.629\ldots$ for $\delta=0$. The limiting situation is special in 
the second case since the resulting spacetime is no longer 
asymptotically flat and since the axis is singular. The invariant 
circumference of the disk is zero in this case which implies that the 
disk shrinks to a point for an observer in the exterior of the disk, 
see \cite{prd4}.

For physical reasons the solution was discussed in~\cite{prd4} in
dependence of two other real parameters $\epsilon$ and $\gamma$. Here
$\epsilon$ is related to the redshift of photons emitted at the center
of the disk and detected at infinity. It varies between 0 in the
Newtonian limit, and 1 in the ultra-relativistic limit, where photons
cannot escape to infinity. Thus, $\epsilon$ is a measure of how
relativistic the situation is. The parameter $\gamma$ is a measure of
how static the solution is, it varies between 0, indicating the static
limit and 1. For the functional relations between $\epsilon$, $\gamma$
and $\lambda$, $\delta$ see~\cite{prd4}. The constant $\Omega$ (with
respect to the physical coordinates) to appear in the following can be
considered as a natural scale for the angular velocities in the disk,
for a definition see~\cite{prd4}.

The coordinate $\rho$ can take all non-negative real values, the
coordinate $\zeta$ all real values. The example we are studying here
has an equatorial symmetry,
\begin{equation}
    \mathcal{E}(\rho,-\zeta)=\bar{\mathcal{E}}(\rho,\zeta)
    \label{eq:eqsym}.
\end{equation}
It is therefore sufficient to consider only non-negative values of
$\zeta$. The case $\rho=0$ corresponds to the axis of symmetry where
the branch cut $[\xi,\bar{\xi}]$ degenerates to a point. As was shown
in~\cite{prd2,prd4}, the Ernst potential can be written in this limit in
terms of theta-functions on the elliptic surface $\mathcal{L}_{1}$
defined by $\mu_{1}^{2}= (K^{2}-\alpha)^{2}+\beta^{2}$, i.e.\ the
surface $\mathcal{L}$ with the cut $[\xi,\bar{\xi}]$ removed. Near the
axis the Ernst potential has the form (see~\cite{fay,prd2})
\begin{equation}
    \mathcal{E}(\rho,\zeta)=\mathcal{E}_{0}(\zeta)+\rho^{2}
    \mathcal{E}_{1}(\zeta)+\mathcal{O}(\rho^{4});
    \label{eq:nearaxis}
\end{equation}
here $\mathcal{E}_{0}$ and $\mathcal{E}_{1}$ are independent of
$\rho$, $\mathcal{E}_{0}$ is the axis potential. This formula could be 
used to calculate the potential close to the axis. However we 
considered only values of $\rho$ greater than
$10^{-5}$ and did not experience any numerical problems. Consequently 
we did not use formula~(\ref{eq:nearaxis}).

For large values of $r=|\xi|$, the Ernst potential has the asymptotic
expansion
\begin{equation}
    \mathcal{E}=1-\frac{2m}{r}+\frac{2m^{2}}{r^{2}}
    -\frac{2iJ\zeta}{r^{3}}+\mathcal{O}(1/r^{3});
    \label{eq:ernstinfinity}
\end{equation}
here the constants (with respect to $\xi$) $m$ and $J$ are the
ADM-mass and, respectively, the angular momentum of the
space-time. They can be calculated on the axis in terms of elliptic
theta-functions, see~\cite{prd4}. Formula~(\ref{eq:ernstinfinity}) is
used for values of $r>10^{6}$.

In the limit $\xi=E_{2}$, the Ernst potential can be given on the
surface $\Sigma_{0}$ of genus~0 obtained by removing the cuts
$[\xi,\bar{\xi}]$ and $[E_{2},F_{2}]$ from the surface
$\mathcal{L}$. The potential can thus be given in this case in terms
of elementary functions, see~\cite{prd4}.

In the equatorial plane $\zeta=0$, the Riemann surface $\mathcal{L}$
has an additional involution $K\to-K$ as can be seen
from~(\ref{hyper1}). This implies that the surface can be considered
as a covering of an elliptic surface, see~\cite{algebro,prd2}. The
theta-functions in~(\ref{ernst2}) can be written as sums of
theta-functions on the covered surface and on the Prym variety which
happens to be an elliptic surface as well in this case. We use this
fact at the disk ($\zeta=0$, $\rho\leq 1$), where the moving branch
points are situated on $\Gamma$. There all quantities can be expressed
in terms of quantities defined on the Prym surface $\Sigma_{w}$
defined by $\mu_{w}^{2}=(K+\rho^{2})((K-\alpha)^{2}+\beta^{2})$,
see~\cite{prd4}.

\section{Numerical implementations}
\label{sec:spectral}

The numerical task in this work is to approximate and evaluate
analytically defined functions as accurately and efficiently as
possible. To this end it is advantageous to use (pseudo-)spectral
methods which are distinguished by their excellent approximation
properties when applied to smooth functions. Here the functions are
known to be analytic except for isolated points. In
this section we explain the basic ideas behind the use of spectral
methods and describe in detail how the theta-functions and the Ernst
potential can be obtained to a high degree of accuracy.

\subsection{Spectral approximation}

The basic idea of spectral methods is to approximate a given function
$f$ globally on its domain of definition by a linear combination
\[
f \approx \sum_{k=0}^N a_k \phi_k,
\]
where the function $\phi_k$ are taken from some class of functions
which is chosen appropriately for the problem at hand.

The coefficients $a_k$ are determined by requiring that the linear
combination should be `close' to $f$. Thus, one could require that
$||f - \sum_{k=0}^N a_k \phi_k||$ should be minimal for some
norm. Another possibility is to require that $\left< f -\sum_{k=0}^N
a_k \phi_k, \chi_l\right> = 0$ for $l=0:N$ with an appropriate inner
product and associated orthonormal basis $\chi_l$. This is called the
Galerkin method. Finally, one can demand that $f(x_l) = \sum_{k=0}^N a_k
\phi_k(x_l)$ at selected points $(x_l)_{l=0:N}$. This is the so
called collocation method which is the one we will use in this
paper. In this case the function values $f_l=f(x_l)$ and the
coefficients $a_k$ are related by the matrix $\Phi_{lk} =
\phi_k(x_l)$.

The choice of the expansion basis depends to a large extent on the
specific problem. For periodic functions there is the obvious choice
of trigonometric polynomials $\phi_k(x) = \exp(2\pi i k/N)$ while for
functions defined on a finite interval the most used functions are
orthogonal polynomials, in particular Chebyshev and Legendre
polynomials. While the latter are important because of their
relationship with the spherical harmonics on the sphere, the former
are used because they have very good approximation properties and
because one can use fast transform methods when computing the
expansion coefficients from the function values provided one chooses
the collocation points $x_l=\cos(\pi l/N)$ (see~\cite{fornberg} and
references therein). We will use here collocation with Chebyshev
polynomials.

Let us briefly summarize here their basic properties. The Chebyshev
polynomials $T_n(x)$ are defined on the interval $I=[-1,1]$ by the
relation 
\[
T_n(\cos(t)) = \cos(n t), \text{where } x = \cos(t),\qquad t\in[0,\pi].
\]
They satisfy the differential equation
\begin{equation}
  \label{eq:diffeqcheb}
  (1-x^2)\, \phi''(x) - x \phi'(x) + n^2 \phi(x) = 0.
\end{equation}
The addition theorems for sine and cosine imply the recursion
relations
\begin{equation}
  \label{eq:recurscheb}
  T_{n+1}(x) - 2 x\, T_n(x) + T_{n-1}(x) = 0, 
\end{equation}
for the polynomials $T_n$ and
\begin{equation}
  \label{eq:recursderiv}
  \frac{T'_{n+1}(x)}{n+1} - \frac{T'_{n-1}(x)}{n-1} = 2 T_n(x)
\end{equation}
for their derivatives. The Chebyshev polynomials are orthogonal on $I$
with respect to the hermitian inner product
\[
\left< f, g \right> = \int_{-1}^1 f(x) \bar g(x) \,\frac{d x}{\sqrt{1-x^2}}.
\]
We have
\begin{equation}
  \label{eq:ortho}
  \left< T_m , T_n \right> = c_m \frac\pi2\, \delta_{mn}
\end{equation}
where $c_0=2$ and $c_l=1$ otherwise.

Now suppose that a function $f$ on $I$ is sampled at the points
$x_l=\cos(\pi l/N)$ and that $\sum_{n=0}^N a_n T_n$ is the
interpolating polynomial. Defining $c_0=c_N=2$, $c_n=1$ for
$0<n<N$ in the discrete case and the numbers $F_n = c_n a_n$ we have
\[
\begin{split}
  f_l &= \sum_{n=0}^N a_n T_n(x_l) = \sum_{n=0}^N a_n T_n(\cos(\pi
  l/N)) \\
  &= \sum_{n=0}^N a_n \cos(\pi nl/N) 
  = \sum_{n=0}^N \frac{F_n}{c_n}\cos(\pi nl/N)
\end{split}.
\]
This looks very much like a discrete cosine series and in
fact one can show~\cite{briggshenson} that the coefficients $F_n$ are
related to the values $f_l$ of the function by an inverse discrete
Fourier transform (DCT)
\[
F_n = \frac2N\sum_{l=0}^N \frac{f_l}{c_l}\cos(\pi nl/N).  
\]
Note, that up to a numerical factor the DCT is idempotent, i.e., it is
its own inverse.  This relationship between the Chebyshev polynomials
and the DCT is the basis for the efficient computations because the
DCT can be performed numerically by using the fast Fourier transform
(FFT) and pre- and postprocessing of the
coefficients~\cite{fornberg}. The fast transform allows us to switch
easily between the representations of the function in terms of its
sampled values and in terms of the expansion coefficients $a_n$ (or
$F_n$).

The fact that $f$ is approximated globally by a finite sum of
polynomials allows us to express any operation applied to $f$
approximately in terms of the coefficients. Let us illustrate this in
the case of integration. So we assume that $f = p_N =\sum_{n=0}^N a_n
T_n$ and we want to find an approximation of the integral for $p_N$,
i.e., the function
\[
F(x) = \int_{-1}^x f(s)\, ds,
\]
so that $F'(x)=f(x)$. We make the ansatz $F(x) = \sum_{n=0}^N b_n\,
T_n(x)$ and obtain the equation
\[
F' = \sum_{n=0}^N b_n\,T'_n = \sum_{n=0}^N a_n T_n = f.
\]
Expressing $T_n$ in terms of the $T'_n$ using~\eqref{eq:recursderiv}
and comparing coefficients implies the equations
\[
b_1 = \frac{2a_{0} - a_{2}}{2}, \qquad b_n = \frac{a_{n-1} -
  a_{n+1}}{2n}\quad \text{for }0< n < N,\qquad b_N = \frac{a_{N-1}}{2N}.
\]
between the coefficients which determines all $b_l$ in terms of the
$a_n$ except for $b_0$. This free constant is determined by the
requirement that $F(-1)=0$ which implies (because $T_n(-1)=(-1)^n$)
\[
b_0 = - \sum_{n=1}^N (-1)^n b_n.
\]
These coefficients $b_n$ determine a polynomial $q_N$ of degree $N$
which approximates the indefinite integral $F(x)$ of the $N$-th degree
polynomial $f$. The exact function is a polynomial of degree $N+1$
whose highest coefficient is proportional to the highest coefficient
$a_N$ of $f$. Thus, ignoring this term we make an error whose
magnitude is of the order of $|a_N|$ so that the approximation will be
the better the smaller $|a_N|$ is. The same is true when a smooth
function $f$ is approximated by a polynomial~$p_N$. Then, again, the
indefinite integral will be approximated well by the polynomial $q_N$
whose coefficients are determined as above provided the highest
coefficients in the approximating polynomial $p_N$ are small.

From the coefficients $b_n$ we can also find an approximation to the
definite integral $\int_{-1}^1 f(s)\,ds = F(1)$ by evaluating 
\[
q_N(1) = \sum_{n=0}^Nb_n = 2\sum_{l=0}^{\lfloor N/2\rfloor}b_{2l+1}.
\]
Thus, to find an approximation of the integral of a function $f$ we
proceed as described above, first computing the coefficients $a_n$ of
$f$, computing the $b_n$ and then calculating the sum of the odd
coefficients. 

\subsection{Implementation of the square-root}

\begin{figure}[htb]
    \centering \epsfig{file=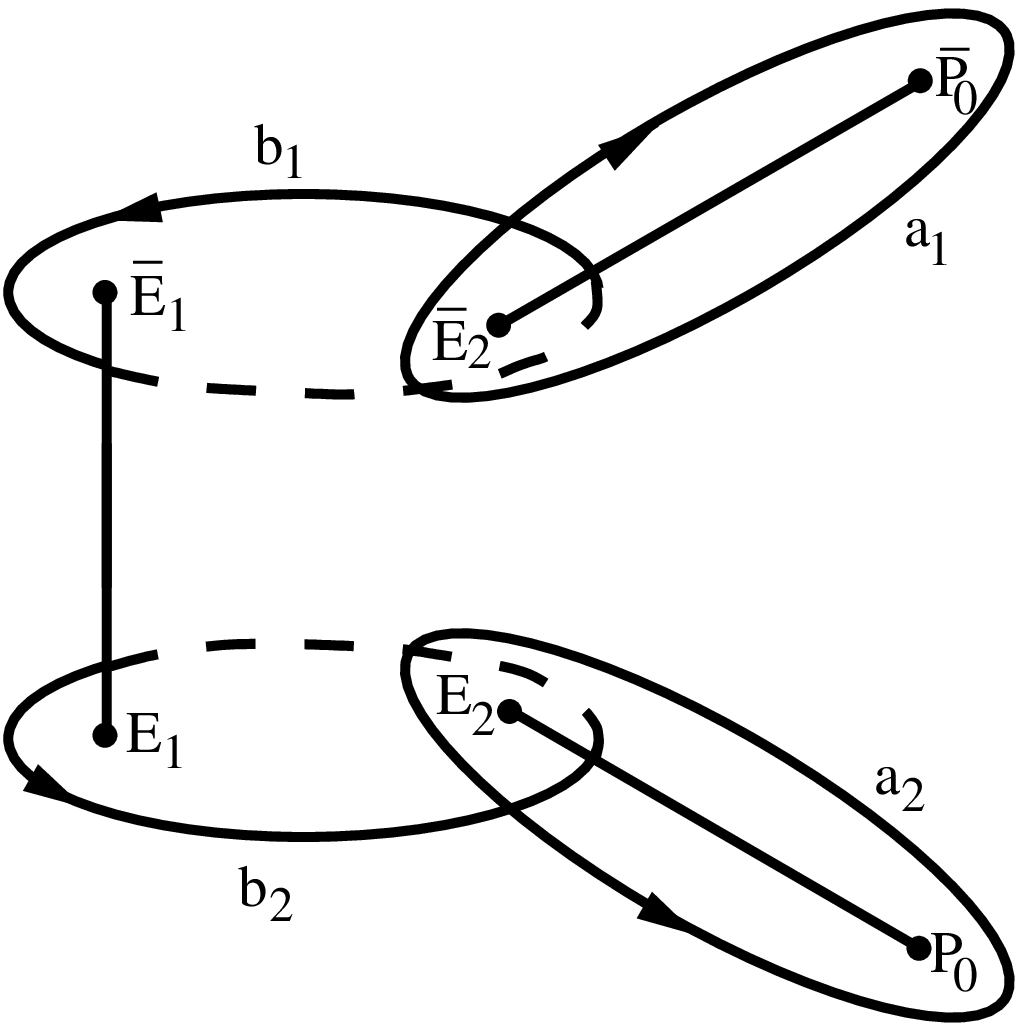,width=6cm}
    \caption{Canonical cycles ($P_{0}=\xi$).}
    \label{fig:cut-system}
\end{figure}

The Riemann surface $\mathcal{L}$ is defined by an algebraic curve of
the form
\[
\mu^{2}=(K-\xi)(K-\bar{\xi})\prod_{i=1}^{g}(K-E_{i})(K-\bar E_{i}),
\]
where in our case we have $g=2$ throughout.  In order to compute the
periods and the theta-functions related to this Riemann surface it is
necessary to evaluate the square-root $\sqrt{\mu^2(K)}$
for arbitrary complex numbers $K$. In order to make this a well
defined problem we introduce the cut-system as indicated in
Fig.~\ref{fig:cut-system}. 
On the cut surface the square-root $\mu(K)$
is defined as in~\cite{heil} as the product of square-roots of
monomials
\begin{equation}
    \mu=\sqrt{K-\xi\phantom{\bar\xi}} \sqrt{K-\bar{\xi}}
    \prod_{i=1}^{g} \sqrt{K-E_{i}} \sqrt{K-\bar E_{i}}.
    \label{eq:root}
\end{equation}
The square-root routines such as the one available in MATLAB usually
have their branch-cut along the negative real axis. The
expression~(\ref{eq:root}) is holomorphic on the cut 
surface so that we cannot simply take the builtin square-root when
computing $\sqrt{\mu^2(K)}$. Instead we need to use the information
provided by the cut-system to define adapted square-roots.

Let $\arg(z)$ be the argument of a complex number $z$ with values in
$]-\pi,\pi[$ and consider two factors in~(\ref{eq:root}) such as 
\[
\sqrt{K-P_1}\sqrt{K-P_2}
\]
where $P_1$ and $P_2$ are two branch-points connected by a
branch-cut. Let $\alpha=\arg(P_2-P_1)$ be the argument at the line
from $P_1$ to $P_2$. Now we define the square-root
$\sqrt[(\alpha)]{\cdot}$ with branch-cut along the ray with argument
$\alpha$ by computing for each $z\in\mathbb{C}$ the square-root
$s:=\sqrt{z}$ with the available MATLAB routine and then putting
\[
\sqrt[(\alpha)]{z} = \left\{
  \begin{array}{rl}
     s & \alpha/2<\arg(s)<\alpha/2 + \pi\\ 
    -s & \text{otherwise}
  \end{array}
\right. .
\]
With this square-root we compute the two factors
\[
\sqrt[(\alpha)]{K-P_1}\sqrt[(\alpha)]{K-P_2}.
\]
It is easy to see that this expression changes sign exactly when the
branch-cut between $P_1$ and $P_2$ is crossed. We compute the
expression~(\ref{eq:root}) by multiplying the pairs of factors which
correspond to the branch-cuts.

This procedure is not possible in the case of the non-linear
transformations we are using to evaluate the periods in certain
limiting cases. In these cases the root is chosen in a way that the
integrand is a continuous function on the path of integration.

\subsection{Numerical treatment of the periods}

The quantities entering formula~(\ref{ernst2}) for the Ernst potential
are the periods of the Riemann surface and the line integrals
$\mathbf{u}$ and $I$. The value of the theta-function is then
approximated by a finite sum.

The periods of a hyperelliptic Riemann surface can be expressed as
integrals between branch points. Since we need in our example the
periods of the holomorphic differentials and the differential of the
third kind with poles at $\infty^{\pm}$, we have to consider integrals
of the form
\begin{equation}
    \int_{P_{i}}^{P_{j}}\frac{K^{n}dK}{\mu(K)}, \quad n=0,1,2
    \label{period1},
\end{equation}
where the $P_{i}$, $i,j=1,\ldots,6$ denote the branch points of
$\mathcal{L}$.

In general position we use a linear transformation of the form $K
=ct+d$ to transform the integral~(\ref{period1}) to the normal form
\begin{equation}
\label{eq:int_aperiod} \int_{-1}^1 \frac{\alpha_0 + \alpha_1 t +
\alpha_2 t^2}{\sqrt{1-t^2}} \;H(t) \,dt,
\end{equation}
where the $\alpha_i$ are complex constants and where $H(t)$ is a
continuous (in fact, analytic) complex valued function on the interval
$[-1,1]$.  This form of the integral suggests to express the powers
$t^n$ in the numerator in terms of the first three Chebyshev
polynomials $T_0(t)=1$, $T_1(t)=t$ and $T_2(t)= 2t^2-1$ and to
approximate the function $H(t)$ by a linear combination of Chebyshev
polynomials 
\[ 
H(t) = \sum_{n\ge0} h_n T_n(t).  
\] 
The integral is then calculated with the help of the orthogonality
relation~(\ref{eq:ortho}) of the Chebyshev polynomials.

Since the Ernst potential has to be calculated for all $\rho,\zeta\in
\mathbb{R}^{+}_{0}$, it is convenient to use the
cut-system~(\ref{fig:cut-system}). In this system the moving cut does
not cross the immovable cut. In addition the system is adapted to the
symmetries and reality properties of $\mathcal{L}$. Thus the periods
$a_{2}$ and $b_{2}$ are related to $a_{1}$ and $b_{1}$ via complex
conjugation. For the analytical calculations of the Ernst potential in
the limit of collapsing cuts, we have chosen in~\cite{prd2} cut
systems adapted to the respective situation. In the limit $\xi\to
\bar{\xi}$ we were using for instance a system where $a_{2}$ is the
cycle around the cut $[\xi,\bar{\xi}]$. This has the effect that only 
the $b$-period $b_{2}$ diverges 
logarithmically in this case whereas the remaining periods stay 
finite as $\rho$ tends to 0. In the cut systems \ref{fig:cut-system}, 
all periods diverge as $\ln \rho$. Since the divergence is only 
logarithmical this does not pose a problem for values of 
$\rho>10^{-5}$. In addition the integrals which have to be calculated 
in the evaluation of the periods are 
the same in both cut-system. Thus  there is no advantage in using 
different cut systems for the numerical work. 

\begin{figure}[htb]
    \centering \epsfig{file=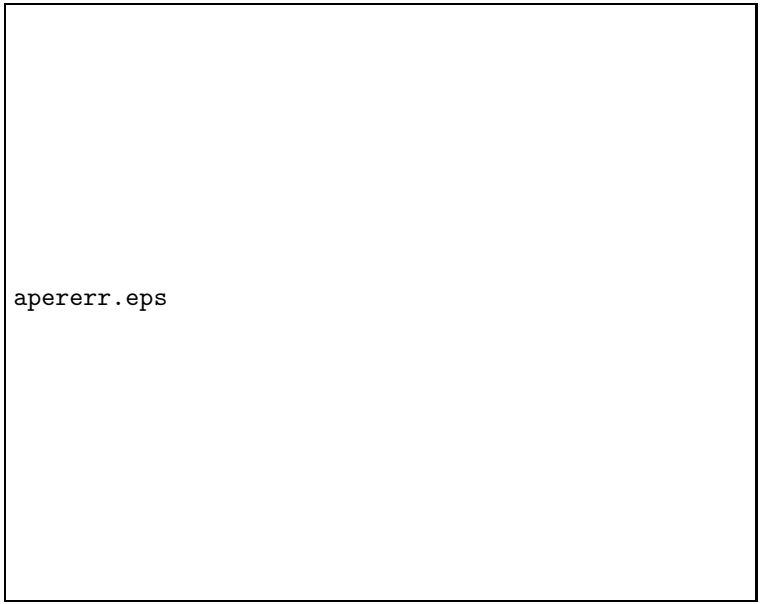,width=10cm}
    \caption{Test of the numerics for the $a$-periods at several 
    points in the space-time. The error is shown in dependence of the 
    number $N$ of Chebychev polynomials.}
    \label{fig:test_periods}
\end{figure}

To test the numerics we use the fact that the integral of any
holomorphic differential along a contour surrounding the cut 
$[E_{1},F_{1}]$ in positive direction is equal to minus the sum of 
all $a$-periods of this integral. Since this condition is not implemented in
the code it provides a strong test for the numerics. It can be seen in
Fig.~\ref{fig:test_periods} that 16 to 32 polynomials are sufficient 
in general position
to achieve optimal accuracy. Since MATLAB works with 16 digits,
machine precision is in general limited to 14 digits due to rounding
errors. These rounding errors are also the reason why the accuracy
drops slightly when a higher number of polynomials is used. The use
of a low number of polynomials consequently does not only require less
computational resources but has the additional benefit of reducing the
rounding errors. It is therefore worthwhile to reformulate a problem
if a high number of polynomials would be necessary to obtain optimal
accuracy.
These situations occur in the calculation of the periods when the moving
branch points almost coincide which happens on the axis of
symmetry in the space-time or at spatial infinity. As can be seen from
Fig.~\ref{fig:test_periods}, for $\rho=10^{-3}$ and $\zeta=10^{3}$ not
even 2048 polynomials (this is the limit due to memory on the low end
computers we were using) produce sufficient accuracy. The reason for
these problems is that the function $H$ in~(\ref{eq:int_aperiod})
behaves like $1/\sqrt{t+\rho}$ near $t=0$. For small $\rho$ this
behavior is only satisfactorily approximated by a large number of
polynomials. We therefore split the integral in two integrals between
$F_{2}$ and $(F_{2}+\bar{\xi})/2$ and between $(F_{2}+\bar{\xi})/2$
and $\bar{\xi}$. The first integral is calculated with the Chebyshev
integration routine after the substitution $t=\sqrt{K-F_{2}}$. This
substitution leads to a regular integrand also at the branch point
$F_{2}$. The second integral is calculated with the Chebyshev
integration routine after the substitution
$K-\zeta=\rho\sinh(t)$. This takes care of the almost collapsing cut
$[\xi,\bar{\xi}]$. It can be seen in Fig.~\ref{fig:test_periods} that
128 polynomials are sufficient to obtain machine precision even in
almost degenerate situations.

\begin{figure}[htb]
    \centering \epsfig{file=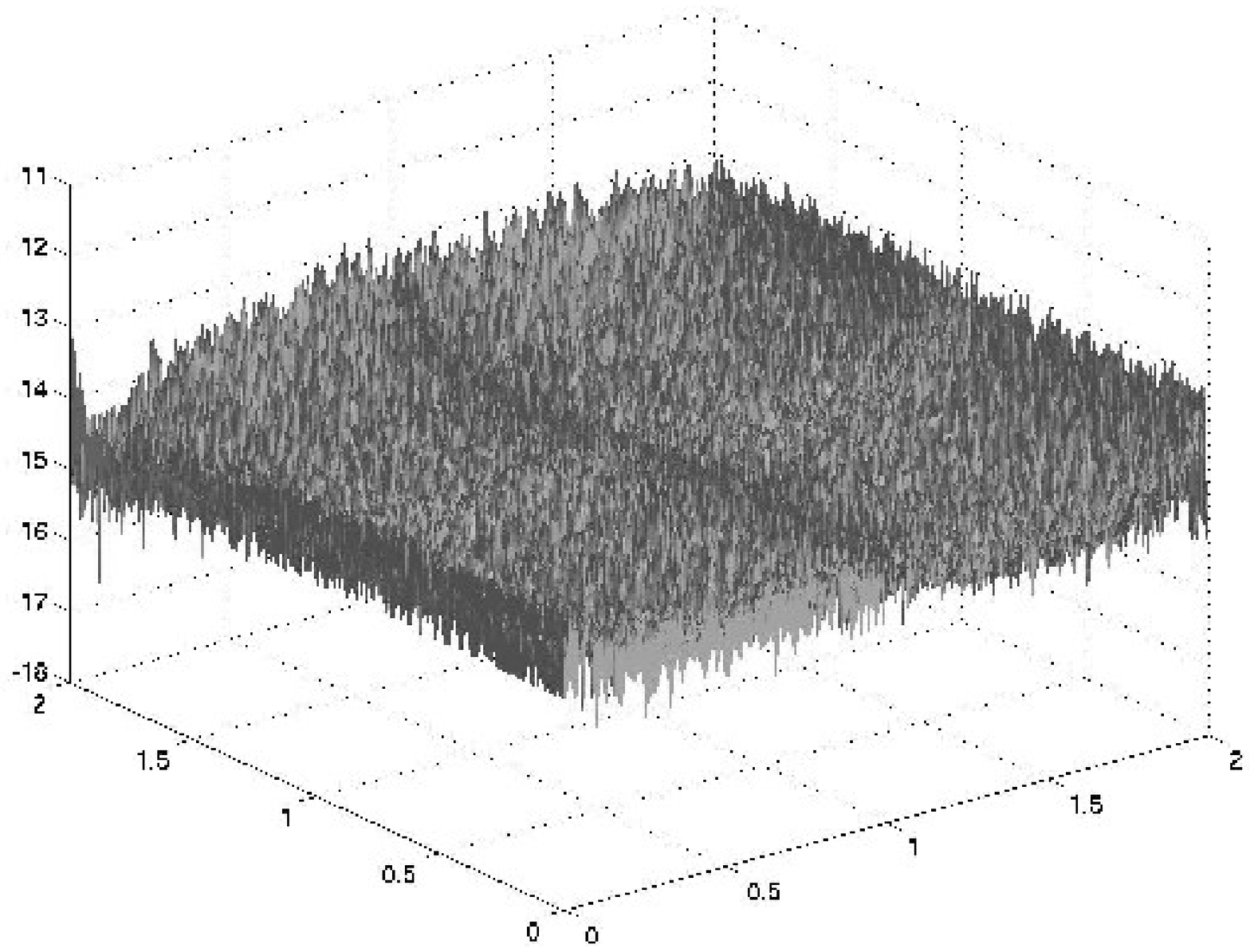,width=10cm}
    \caption{A measure for the error in the determination of the 
    periods in dependence of the physical coordinates. 
    For $\rho,\zeta>1$ we use $1/\rho,1/\zeta$ as
    coordinates.}
    \label{fig:error}
\end{figure}

The cut-system in Fig.~\ref{fig:cut-system} is adapted to the limit
$\bar{\xi}\to F_{2}$ in what concerns the $a$-periods, since the cut
which collapses in this limit is encircled by an $a$-cycle. However
there will be similar problems as above in the determination of the
$b$-periods. For $\bar{\xi}\sim F_{2}$ we split the integrals for the
$b$-periods as above in two integrals between $F_{1}$ and $0$, and $0$
and $F_{2}$. For the first integral we use the integration variable $t
= \sqrt{K-F_1}$, for the second $K=\Re F_{2}-i\Im F_{2}\sinh t$. Since
the Riemann matrix (the matrix of $b$-periods of the holomorphic
differentials after normalization) is symmetric, the error in the
numerical evaluation of the $b$-periods can be estimated via the
asymmetry of the calculated Riemann matrix. We define the function
$err(\rho,\zeta)$ as the maximum of the norm of the difference in the
$a$-periods discussed above and the difference of the off-diagonal
elements of the Riemann matrix. This error is presented for a whole
space-time in Fig.~\ref{fig:error}. The values for $\rho$ and $\zeta$
vary between $10^{-4}$ and $10^{4}$. On the axis and at the disk we
give the error for the elliptic integrals (only the error in the 
evaluation of the $a$-periods, since the
Riemann matrix has just one component). For $\xi\to \infty$ the
asymptotic formulas for the Ernst potential are used. The calculation
is performed with 128 polynomials, and up to 256 for
$|\xi|>10^{3}$. It can be seen that the error is in this case globally
below $10^{-13}$.

\subsection{Numerical treatment of the line integrals}
\begin{figure}[htb]
    \centering \epsfig{file=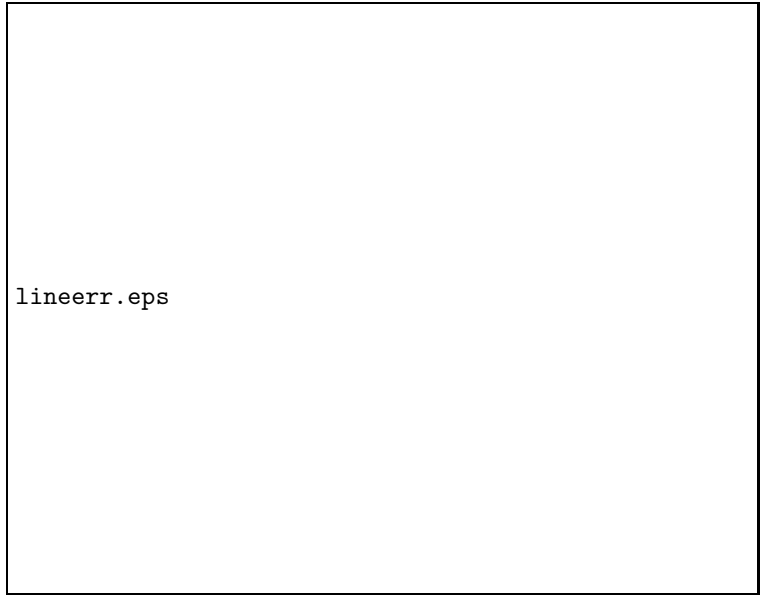,width=10cm}
    \caption{Error in the integrals for the Maclaurin solution in 
    dependence of the number $N$ of Chebychev polynomials.}
    \label{fig:line}
\end{figure}

The line integrals $\mathbf{u}$ and $I$ in~(\ref{ernst2}) are linear
combinations of integrals of the form
\begin{equation}
    \int_{-i}^{i}\frac{\ln G(K)K^{l}dK}{\mu(K)}, \qquad l=0,1,2
    \label{eq:line1}.
\end{equation}
In general position, i.e.\ not close to the disk and $\lambda$ small
enough, the integrals can be directly calculated after the
transformation $K=it$ with the Chebyshev integration routine. To test
the numerics we consider the Newtonian limit ($\lambda\to0$) where the
function $\ln G$ is proportional to $1+K^{2}$, i.e.\ we calculate the
test integral
\begin{equation}
    \int_{-i}^{i}\frac{(1+K^{2})\;dK}{\sqrt{(K-\zeta)^{2}+\rho^{2}}}
    \label{eq:testline}.
\end{equation}
 We compare the numerical with the analytical result in
 Fig.~\ref{fig:line}. In general position machine precision is reached
 with 32 polynomials.

When the moving cut approaches the path $\Gamma$, i.e., when the
space-time point comes close to the disk, the integrand
in~(\ref{eq:testline}) develops cusps near the points $\xi$ and
$\bar{\xi}$. In this case a satisfactory approximation becomes
difficult even with a large number of polynomials. 
Therefore we split the integration path in $[-i,-i\rho]$, 
$[-i\rho,i\rho]$ 
and $[i\rho,i]$. Using the reality properties of the integrands, we 
only calculate the integrals between $0$ and $i\rho$, and between 
$i\rho$ and $i$. In the first case we use the transformation $K= 
\zeta+\rho\sinh t$ to evaluate the integral with the Chebyshev 
integration routine, in the second case we use 
the transformation $t = \sqrt{K-\bar{\xi}}$. 
It can be seen 
in figure \ref{fig:line} that machine precision can be reached even 
at the disk with 64 to 128 polynomials.  The values at the disk are, 
however, determined in terms of elliptic functions which is more 
efficient than the hyperelliptic formulae.

\begin{figure}[htb]
    \centering \epsfig{file=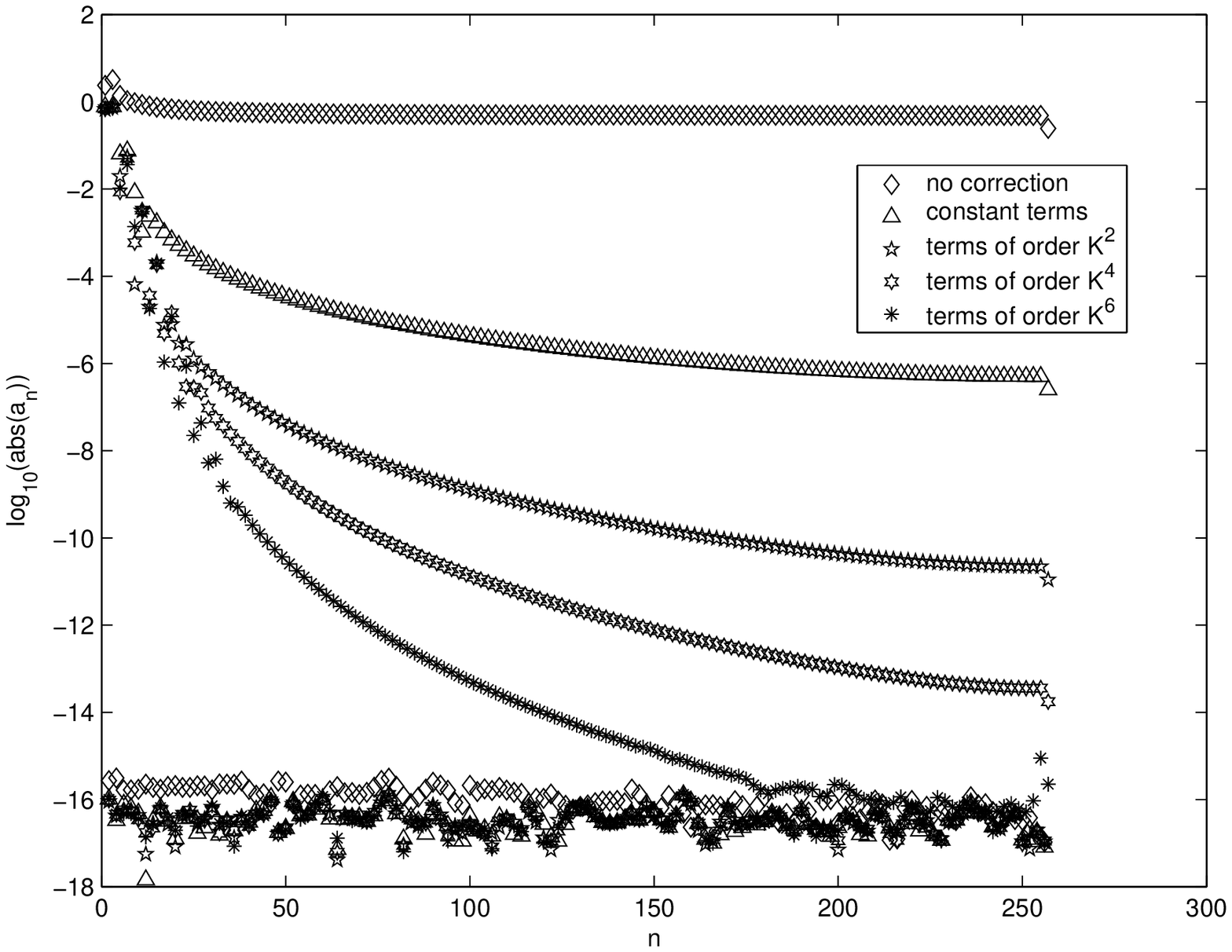,width=10cm}
    \caption{Spectral coefficients for the integral~(\ref{eq:axisreg}) 
    for $\delta=1$ and $\lambda=10^{16}$ in dependence of the number 
    of Chebychev polynomials.}
    \label{fig:logreg}
\end{figure}

To treat the case where $\delta\lambda^{2}$ is not small, it is
convenient to rewrite the function $G$ in~(\ref{disk2}) in the form
\begin{equation}
    \ln G(K) =2\ln \left(\sqrt{(K^{2}-\alpha)^{2}+\beta^{2}}+K^{2}
    +1\right)-\ln \left(\frac{1}{\lambda^{2}}-\delta K^{2}\right)
    \label{eq:log}.
\end{equation}
In the limit $\delta \lambda^{2}\to \infty$ with $\delta$ finite, the
second term in~(\ref{eq:log}) becomes singular for $K=0$. Even for
$\delta\lambda^{2}$ large but finite, the approximation of the
integrand by Chebyshev polynomials requires a huge number of
coefficients as can be seen from Fig.~\ref{fig:logreg}. It is
therefore sensible to `regularize' the integrand near $K=0$. We
consider instead of the function $\ln( \frac{1}{\lambda^{2}}-\delta
K^{2}) F(K)$ where $F(K)$ is a $C^{\infty}$ function near $K=0$, the
function
\begin{equation}
    \ln \left(\frac{1}{\lambda^{2}}-\delta K^{2}\right)\left(
    F(K)-F(0)-F'(0)K-\ldots-\frac{1}{n!}F^{(n)}(0)K^{n}\right)
    \label{eq:logreg}.
\end{equation}
The parameter $n$ is chosen such that the spectral coefficients
of~(\ref{eq:logreg}) are of the order of $10^{-14}$ for a given number
of polynomials, see Fig.~\ref{fig:logreg}. There we consider the
integral
\begin{equation}
    \int_{-i}^{i}\frac{\ln
    G(K)dK}{\sqrt{(K^{2}-\alpha)^{2}+\beta^{2}}}
    \label{eq:axisreg},
\end{equation}
which has to be calculated on the axis. We show the absolute values of
the coefficients $a_{k}$ in an expansion of the integrand in Chebyshev
polynomials, $\sum_{k=1}^{N}a_{k}T_{k}$.  It can be seen that one has
to include values of $n=6$ in~(\ref{eq:logreg}). The integral
$\int_{\Gamma}^{}\ln G(K) F(K)$ is then calculated numerically as the
integral of the function~(\ref{eq:logreg}), the subtracted terms are
integrated analytically.
In this way one can ensure that the line integrals are calculated in
the whole space-time with machine precision: close to the Newtonian
limit, we use an analytically known test function to check the
integration routine, for general situations we check the quality of
the approximation of the integrand by Chebyshev polynomials via the
spectral coefficients which have to become smaller than $10^{-14}$.

\subsection{Theta-functions}

The theta series~(\ref{theta}) for the Riemann theta-function (the 
theta function in (\ref{theta}) with zero characteristic, theta 
functions with characteristic follow from (\ref{theta2})) is
approximated as the sum
\begin{equation}
    \Theta(\mathbf{x}|\mathbf{B})
    =\sum_{n_{1}=-N}^{N}\sum_{n_{2}=-N}^{N}\exp\left\{
    \frac{1}{2}n_{1}^{2}B_{11}+n_{1}n_{2}B_{12}+\frac{1}{2}B_{22}
    +n_{1}x_{1}+n_{2}x_{2}\right\}.
    \label{eq:thetasum}
\end{equation}
The value of $N$ is determined by the condition that terms in the
series~(\ref{theta}) for $n>N$ are strictly smaller than some
threshold value $\epsilon$ which is taken to be of the order of
$10^{-16}$. To this end we determine the eigenvalues of $\mathbf{B}$
and demand that
\begin{equation}
    N> -\frac{1}{B_{max}}\left(||\mathbf{x}||+\sqrt{||\mathbf{x}||^{2}
    +2\ln \epsilon B_{max}}\right)
    \label{eq:N},
\end{equation}
where $B_{max}$ is the real part of the eigenvalue with maximal real
part ($\mathbf{B}$ is negative definite). For a more sophisticated
analysis of theta summations see~\cite{deconinck03}. In general
position we find values of $N$ between 4 and 8. For very large values
of $\zeta$ close to the axis, $N$ can become larger that 40 which
however did not lead to any computational problems. To treat more
extreme cases it could be helpful to take care of the fact that the
eigenvalues of $\mathbf{B}$ can differ by more than an order of
magnitude in our example. In these cases a summation over an ellipse
rather than over a sphere in the plane $(n_{1},n_{2})$, i.e.\
different limiting values for $n_{1}$ and $n_{2}$ as
in~\cite{deconinck03} will be more efficient. 

\begin{figure}[htb]
    \centering \epsfig{file=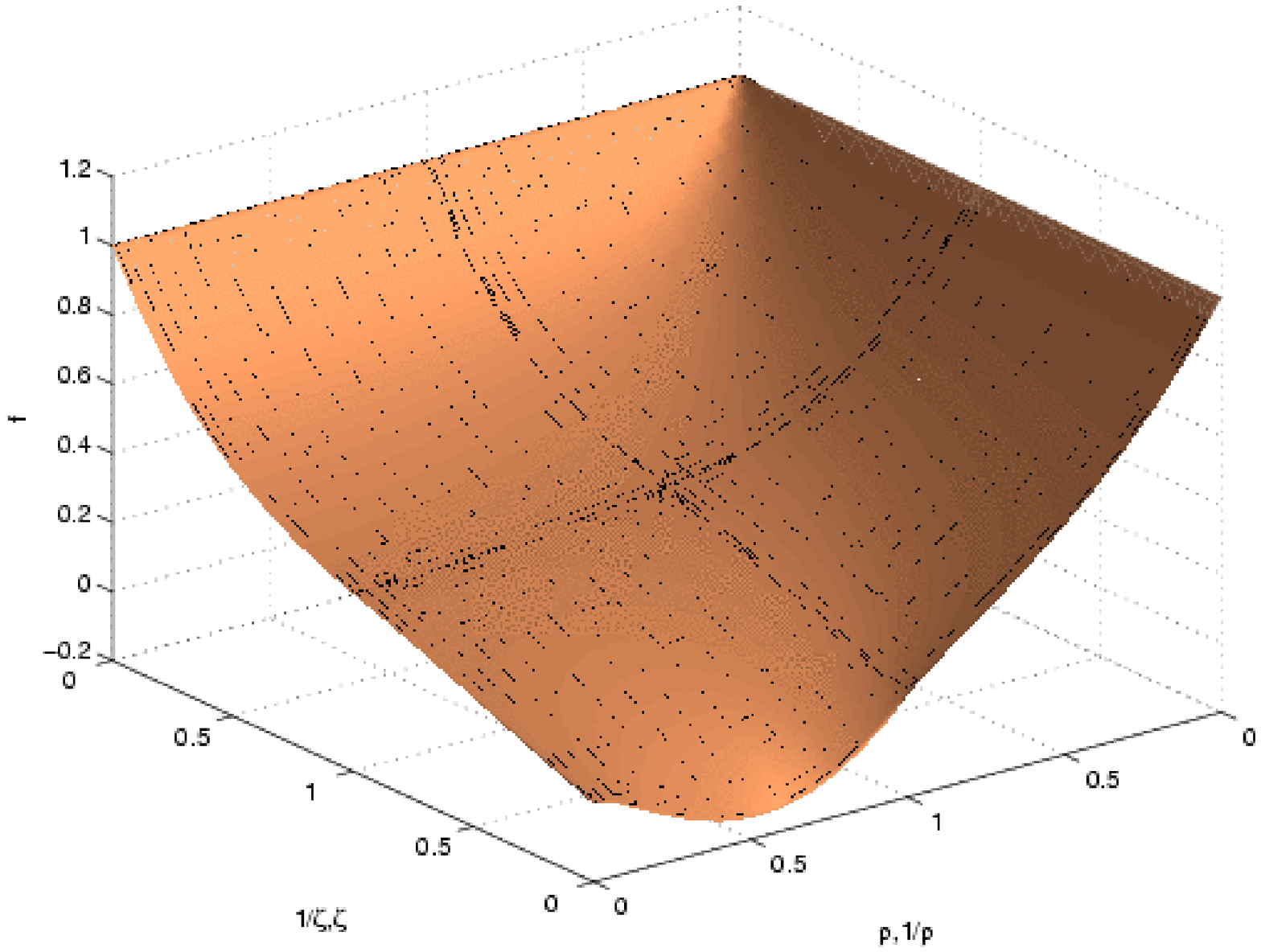,width=10cm}
    \caption{The real part of the Ernst potential for $\lambda=10$ and 
    $\delta=1$ in dependence of the physical coordinates. 
    For $\rho,\zeta>1$ we use $1/\rho,1/\zeta$ as 
    coordinates.}
    \label{fig:f}
\end{figure}

In our case the computation
of the integrals entering the theta-functions was however always the
most time consuming such that an optimization of the summation of the
theta-function would not have a noticeable effect. Due to the 
vectorization techniques in MATLAB, the theta summation always took 
less than 10 \% of the calculation time for a value of the Ernst 
potential. Between 50 and 70 \% of the processor time are used for the 
determination of the periods. On the used low-end PCs, the calculation 
time varied between 0.4 and 1.2s depending on the used number of 
polynomials. 
We show a plot of the real part of the Ernst potential for
$\lambda=10$ and $\delta=1$ in Fig.~\ref{fig:f}. For $\rho,\zeta>1$,
we use $1/\rho,1/\zeta$ as coordinates which makes it possible to plot
the whole space-time in Weyl coordinates. The non-smoothness of the
coordinates across $\rho=1=1/\rho$ and $\zeta=1=1/\zeta$ is noticeable
in the plot. Asymptotically the potential is equal to 1. The disk is
situated in the equatorial plane between $\rho=0$ and $\rho=1$. At the
disk, the normal derivatives of $f$ are discontinuous.

\begin{figure}[htb]
    \centering \epsfig{file=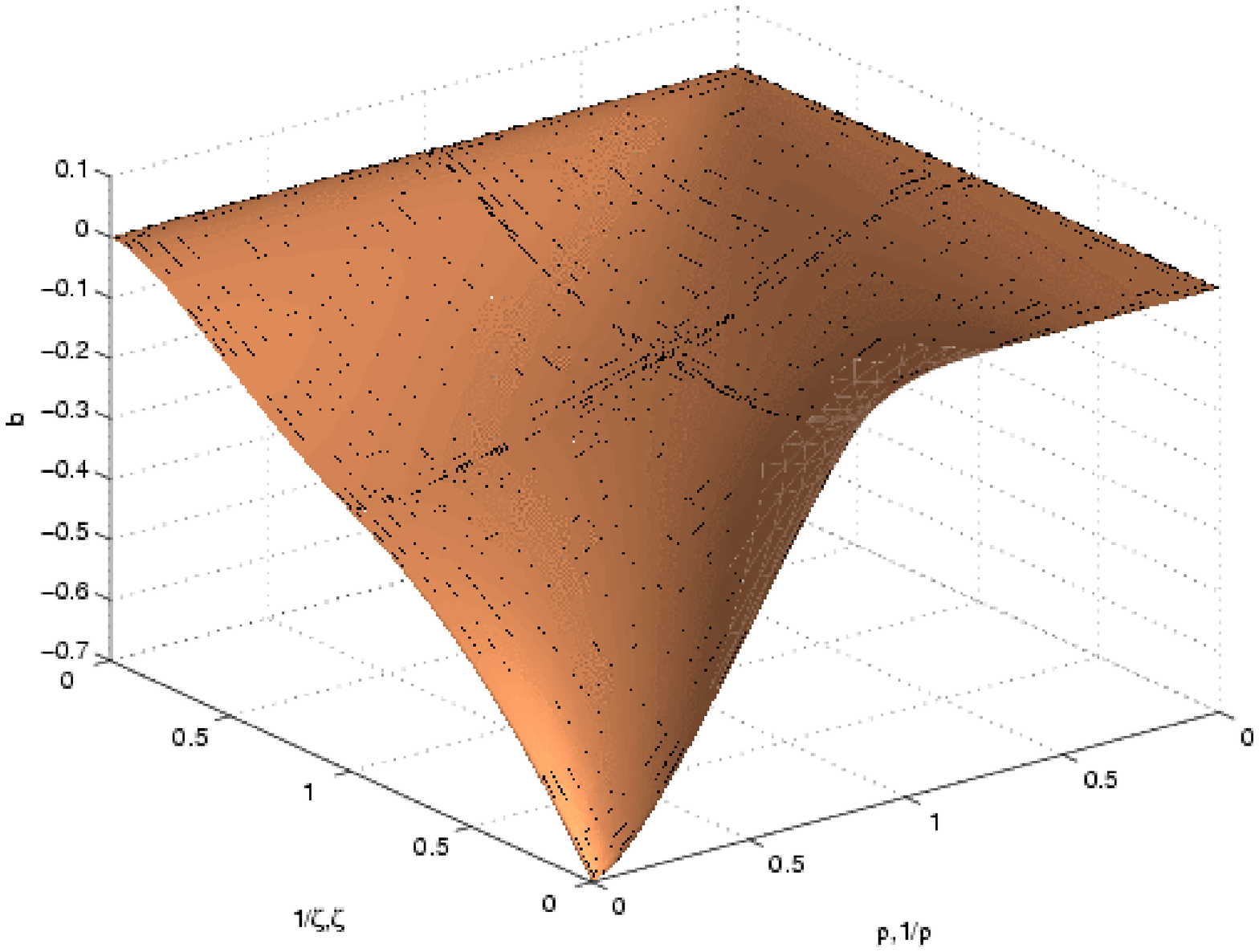,width=10cm}
    \caption{The imaginary part of the Ernst potential for $\lambda=10$ and 
    $\delta=1$ in dependence of the physical coordinates. 
    For $\rho,\zeta>1$ we use $1/\rho,1/\zeta$ as 
    coordinates.}
    \label{fig:b}
\end{figure}

The imaginary part of the Ernst potential in this case is given in
Fig.~\ref{fig:b}. It vanishes at infinity and at the regular part of
the equatorial plane. At the disk, the potential has a jump.

\section{Integral identities}
\label{sec:integrals}
In the previous section we have tested the accuracy of the numerics
locally, i.e.\ at single points in the space-time. Integral identities
have the advantage that they provide some sort of global test of the
numerical precision since they sum up the errors. In addition they
require the calculation of the potentials in extended regions of the
space-time which allows to explore the numerics for rather general
values of the physical coordinates.

The identities we are considering in the following are the well known
equivalence of a mass calculated at the disk (the Komar mass) and the
ADM mass determined at infinity, see~\cite{komar,wald}, and a generalization
of the Newtonian virial identity, see~\cite{virial} and the
appendix. The derivatives of the Ernst potential occurring in the
integrands can be related to derivatives of theta-functions,
see~\cite{prd2}. Since we are interested here in the numerical
treatment of theta-functions with spectral methods, we determine the
derivatives with spectral methods, too (see section 3). The integrals
are again calculated with the Chebyshev integration routine. The main
problem in this context is the singular behavior of the integrands
e.g.\ at the disk which is a singularity for the space-time. As before
this will lead to problems in the approximation of these terms via
Chebyshev polynomials. This could lead to a drop in accuracy which is
mainly due to numerical errors in the evaluation of the integrand and
not of the potentials which we want to test. An important point is
therefore the use of integration variables which are adapted to the
possible singularities.

\subsection{Mass equalities}
The equality between the ADM mass and the Komar mass provides a test
of the numerical treatment of the elliptic theta-functions at the disk
by means of the elliptic theta-functions on the axis.  Since this
equality is not implemented in the code, it provides a strong test.

The Komar mass at the disk is given by formula~(\ref{virial2}) of the
appendix. In the example we are considering here, the normal
derivatives at the disk can be expressed via tangential derivatives
(see~\cite{prd4}) which makes a calculation of the derivatives solely
within the disk possible. We implement the Komar mass in the
form
\begin{equation}
    m_{K}= \int_{0}^{1}d\rho\frac{b_{\rho}}{4\Omega^{2}\sqrt{\rho^{2}-\delta
    f^{2}+2f/\lambda}}
    \left(f+\frac{\Omega^{2}}{f}(\rho^{2}-a^{2}f^{2})\right)
    \label{eq:komar2}.
\end{equation}

\begin{figure}[htb]
    \centering \epsfig{file=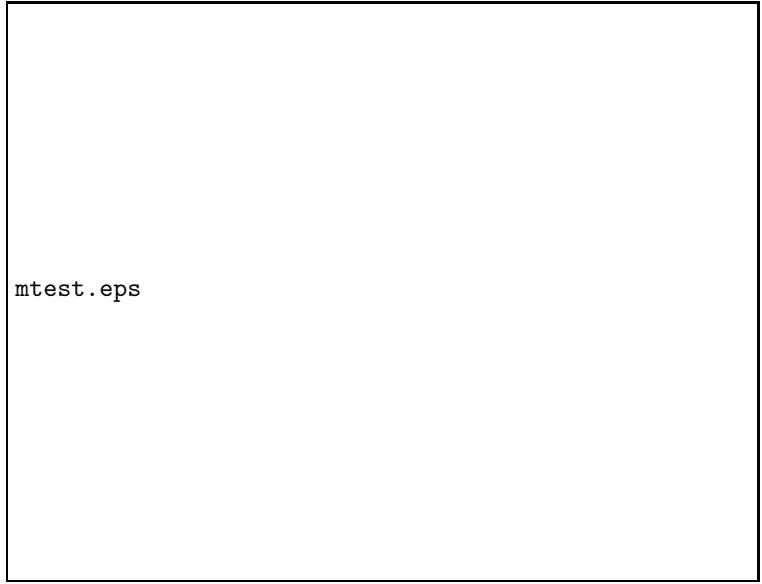,width=10cm}
    \caption{The relative difference of the ADM mass and the Komar 
    mass for $\gamma=0.7$ and several values of $\epsilon$, and for
    $\epsilon=0.8$ and several values of $\gamma$.}
    \label{fig:mtest}
\end{figure}

The integrand is known to vanish as $\sqrt{1-\rho^{2}}$ at the rim of
the disk, which is the typical behavior for such disk solutions. Since
$\sqrt{1-\rho^{2}}$ is not analytic in $\rho$, an expansion of the
integrand~(\ref{eq:komar2}) in Chebyshev polynomials in $\rho$ would
not be efficient. We will thus use $t= \sqrt{1-\rho^{2}}$ as the
integration variable. This takes care of the behavior at the rim of
the disk. Since in general the integrand in \ref{eq:komar2} depends on
$\rho^{2}$, this variable can be used in the whole disk. In the
ultra-relativistic limit for $\delta\neq 0$, the function $f$ vanishes
as $\rho$. In such cases it is convenient either to take two domains
of integration or to use a different variable of integration. We chose
the second approach with $\rho=\sin x$ (this corresponds to the disk 
coordinates (\ref{eq:diskcoor})). Yet, strongly
relativistic situations still lead to problems since $f$ vanishes in this
case at the center of the disk as does $b_{\rho}$ which leads to a
`0/0' limit. In Fig.~\ref{fig:mtest} one can see that the masses
are in general equal to the order of $10^{-14}$. In these
calculations 128 up to 256 polynomials were used. We show the
dependence for $\gamma=0.7$ and several values of $\epsilon$, as well
as for $\epsilon=0.8$ and several values of $\gamma$. The accuracy
drops in the strongly relativistic, almost static situations
($\epsilon$ close to 1, $\gamma$ close to zero) since the Riemann
surface is almost degenerate in this case ($\beta\to 0$). In the
ultra-relativistic limit for $\delta=0$, the situation is no longer
asymptotically flat which implies that the masses formally
diverge. For $\epsilon=0.95$, the masses are still equal to the order
of $10^{-13}$. Not surprisingly the accuracy drops for
$\epsilon=0.9996$ to the order of $10^{-4}$.

\subsection{Virial-type identities}
Generalizations of the Newtonian virial theorem are used in numerics
(see~\cite{virial}) as a test of the quality of the numerical solution
of the Einstein equations. Since they involve integrals over the whole
space-time, they test the numerics globally and thus provide a valid
criterion for the entire range of the physical coordinates.

The identity which is checked here is a variant of the one given
in~\cite{virial} which is adapted to possible problems at the zeros of
the real part of the Ernst potential, the so-called ergosphere,
see~\cite{prd4} for the disk solutions discussed
here. Eq.~(\ref{virial20}) relates integrals of the Ernst
potential and its derivatives over the whole space-time to
corresponding integrals at the disk. Since the numerics at the disk
has been tested above, this provides a global test of the evaluation
of the Ernst potential. As before, derivatives and integrals will be
calculated via spectral methods.

The problem one faces when integrating over the whole space-time is
the singular behavior of the fields on the disk which represents a
discontinuity of the Ernst potential. The Weyl coordinates in which
the solution is given are not optimal to describe the geometry near
the disk. Hence a huge number of polynomials is necessary to
approximate the integrands in~(\ref{virial20}). Even with $512$
polynomials for each coordinate, the coefficients of an expansion in
Chebyshev polynomials did not drop below $10^{-6}$ in more
relativistic situations. Though the computational limits are reached,
the identity~(\ref{virial20}) is only satisfied to the order of
$10^{-8}$ which is clearly related to the bad choice of coordinates.

We therefore use for this calculation so-called disk coordinates
$\eta$, $\theta$ (see~\cite{bitr}) which are related to the Weyl
coordinates via
\begin{equation}
    \rho+i\zeta=\cosh(\eta+i\theta)
    \label{eq:diskcoor}.
\end{equation}
The coordinate $\eta$ varies between $\eta=0$, the disk, and infinity,
the coordinate $\theta$ between $-\pi/2$ and $\pi/2$. The axis is
given by $\pm \pi/2$, the equatorial plane in the exterior of the disk
by $\theta=0$ and $\eta\neq0$. Because of the equatorial symmetry, we
consider only positive values of $\theta$. The surfaces of constant
$\eta$ are confocal ellipsoids which approach the disk for small
$\eta$. For large $\eta$, the coordinates are close to spherical
coordinates.

\begin{figure}[htb]
    \centering
    \epsfig{file=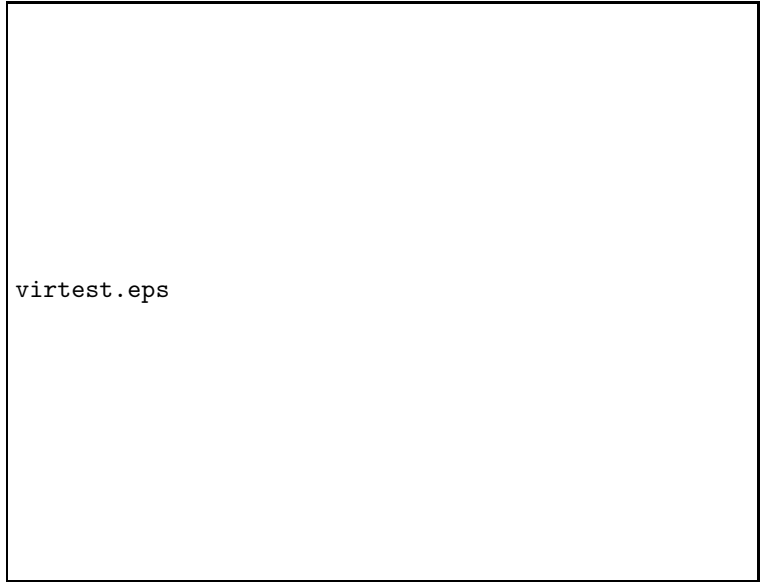,width=10cm}
    \caption{The deviation from zero of the virial-type identity 
    for $\gamma=0.7$ and several values of $\epsilon$, and for
    $\epsilon=0.8$ and several values of $\gamma$.}
    \label{fig:virialtest}
\end{figure}

To evaluate the integrals in~(\ref{virial20}), we perform the
$\eta$-integration up to a value $\eta_{0}$ as well as the
$\theta$-integration with the Chebyshev integration routine. The
parameter $\eta_{0}$ is chosen in a way that the deviation from
spherical coordinates becomes negligible, typically $\eta_{0}=15$. The
integral from $\eta_{0}$ to infinity is then carried out analytically
with the asymptotic formula~(\ref{eq:ernstinfinity}). It turns out
that an expansion in $64$ to $128$ polynomials for each coordinate is
sufficient to provide a numerically optimal approximation within the
used precision. This illustrates the convenience of the disk
coordinates in this context. The virial identity is then satisfied to
the order of $10^{-12}$. We plot the deviation of the sum of the
integrals in~(\ref{virial20}) from zero for several values of
$\lambda$ and $\gamma$ in Fig.~\ref{fig:virialtest}.
The drop in accuracy for strongly relativistic almost static 
situations ($\gamma$ small and $\epsilon$ close to 1) is again due 
to the almost degenerate Riemann surface. The lower accuracy in the 
case of strongly relativistic situations for $\gamma=1$ reflects the 
fact that the disk is shrinking to a point in this limit. To 
maintain the needed resolution one would have to use more polynomials 
in the evaluation of the virial-type identity which was not possible 
on the used computers.

\section{Testing \texttt{LORENE}}
\label{sec:lorene}
One purpose of exact solutions of the Einstein equations is to provide
test-beds for numerical codes to check the quality of the numerical
approximation. In the previous sections we have established that the
theta-functional solutions can be numerically evaluated to the order
of machine precision which implies they can be used in this respect.

The code we are considering here is a C++-library called
\texttt{LORENE}~\cite{Lorene} which was constructed to treat problems from
relativistic astrophysics such as rapidly rotating neutron stars. The
main idea is to solve Poisson-type equations iteratively via spectral
methods. To this end an equation as the Ernst equation~(\ref{ernst1})
is written in the form
\begin{equation}
    \Delta \mathcal{F} = \mathcal{G}(\mathcal{F},r,\theta,\phi)
    \label{eq:poisson},
\end{equation}
where spherical coordinates $r$, $\theta$, $\phi$ are used, and where
$\mathcal{G}$ is some possibly non-linear functional of $\mathcal{F}$
and the coordinates. The system~(\ref{eq:poisson}) is to be solved for
$\mathcal{F}$ which can be a vector. In an iterative approach, the
equation is rewritten as
\begin{equation}
    \Delta \mathcal{F}_{n+1} =
    \mathcal{G}(\mathcal{F}_{n},r,\theta,\phi),\quad n=1,2,\ldots
    \label{eq:poisson2}.
\end{equation}
Starting from some initial function $\mathcal{F}_{0}$, in each step of
the iteration a Poisson equation is solved for a known right-hand
side. For the stationary axisymmetric Einstein equations which we are
considering here, it was shown in~\cite{schaudt} that this iteration
will converge exponentially for small enough boundary data if the
initial values are close to the solution of the equation in some
Banach space norm. It turns out that one can always start the
iteration with Minkowski data, but it is necessary to use a
relaxation: instead of the solution $\mathcal{F}_{n+1}$
of~(\ref{eq:poisson2}), it is better to take a combination
$\tilde{\mathcal{F}}_{n+1}=\mathcal{F}_{n+1}+\kappa \mathcal{F}_{n}$
with $\kappa\in ]0,1[$ (typically $\kappa=0.5$) as a new value in the
source $\mathcal{G}_{n+1}$ to provide numerical stability. The
iteration is in general stopped if $||\mathcal{F}_{n+1}
-\mathcal{F}_{n}||<10^{-10}$.

The Ernst equation~(\ref{ernst1}) is already in the
form~(\ref{eq:poisson}), but it has the disadvantage that the equation
is no longer strongly elliptic at the ergo-sphere where
$\Re(\mathcal{E})=0$. In physical terms, this apparent singularity is 
just a coordinate singularity, and the theta-functional solutions are
analytic there. The Ernst equation in the form~(\ref{eq:poisson}) has
a right-hand side of the form `$0/0$' for $\Re \mathcal{E}=0$ which
causes numerical problems especially in the iteration process since
the zeros of the numerator and the denominator will only coincide for
the exact solution.  The disk solutions we are studying here have
ergo-spheres in the shape of cusped toroids
(see~\cite{prd4}). Therefore it is difficult to take care of the limit
$0/0$ by using adapted coordinates.  Consequently the use of the Ernst
picture is restricted to weakly relativistic situations without
ergo-spheres in this framework.

To be able to treat strongly relativistic situations, we use a
different form of the stationary axisymmetric vacuum Einstein
equations which is derived from the standard $3+1$-decomposition,
see~\cite{eric1}. We introduce the functions $\nu$ and $N_{\phi}$ via
\begin{equation}
    e^{2\nu}=\frac{\rho^{2}f}{\rho^{2}-a^{2}f^{2}},\quad
    N_{\phi}=\frac{\rho af^{2}}{\rho^{2}-a^{2}f^{2}},
    \label{eq:nuN}
\end{equation}
where $ae^{2U}$ is the $g_{t\phi}$ component of the metric leading to
the Ernst potential, see~(\ref{eq:wlp}) in the appendix. Expressions
for $a$ in terms of theta-functions are given in~\cite{prd4}. The
vacuum Einstein equations for the functions~(\ref{eq:nuN}) read
\begin{eqnarray}
    \Delta \nu & = & \frac{1}{2}\rho^{2}e^{-4\nu}(N_{\phi,\rho}^{2}+
    N_{\phi,\zeta}^{2})
    \label{eq:nu},  \\
    \Delta N_{\phi} -\frac{1}{\rho^{2}}N_{\phi}& = &
    4\rho(N_{\phi,\rho}(e^{2\nu})_{\rho}+
    N_{\phi,\zeta}(e^{2\nu})_{\zeta}).
    \label{eq:Npfi}
\end{eqnarray}
By putting $V=N_{\phi}\cos\phi$
we obtain the flat 3-dimensional Laplacian acting on $V$ on the
left-hand side,
\begin{equation}
    \Delta V = 4\rho(V_{\rho}(e^{2\nu})_{\rho}+
    V_{\zeta}(e^{2\nu})_{\zeta}).
    \label{eq:V}
\end{equation}
Since the function $e^{2\nu}$ can only vanish at a horizon, it is
globally non-zero in the examples we are considering here. Thus the
system of equations~(\ref{eq:nu}) and~(\ref{eq:V}) is strongly
elliptic, even at an ergo-sphere.

The disadvantage of this regular system is the non-linear dependence
of the potentials $\nu$ and $N_{\phi}$ on the Ernst potential and $a$
via (\ref{eq:nuN}). Thus we loose accuracy due to rounding errors of
roughly an order of magnitude. Though we have shown in the previous
sections that we can guarantee the numerical accuracy of the data for
$f$ and $af$ to the order of $10^{-14}$, the values for $\nu$ and $V$
are only reliable to the order of $10^{-13}$.

To test the spectral methods implemented in \texttt{LORENE}, we
provide boundary data for the disk solutions discussed above on a
sphere around the disk. For these solutions it would have been more
appropriate to prescribe data at the disk, but \texttt{LORENE} was
developed to treat objects of spherical topology such as stars which
suggests the use of spherical coordinates. It would be possible to
include coordinates like the disk coordinates of the previous section
in \texttt{LORENE}, but this is beyond the scope of this
article. Instead we want to use the Poisson-Dirichlet routine which
solves a Dirichlet boundary value problem for the Poisson equation for
data prescribed at a sphere. We prescribe the data for $\nu$ and
$N_{\phi}$ on a sphere of radius $R$ and solve the
system~(\ref{eq:nu}) and~(\ref{eq:V}) iteratively in the exterior of
the sphere. If the iteration converges, we compare the numerical
solution in the exterior of the sphere with the exact solution.

Since spherical coordinates are not adapted to the disk geometry, a
huge number of spherical harmonics would be necessary to approximate
the potentials if $R$ is close to the disk radius. The limited memory
on the used computers imposes an upper limit of 64 to 128 harmonics. We
choose the radius $R$ and the number of harmonics in a way that the
Fourier coefficients in $\theta$ drop below $10^{-14}$ to make sure
that the provided boundary data contain the related information to the
order of machine precision. The exterior of the sphere where the
boundary data are prescribed is divided in two domains, one from $R$
to $2R$ and one from $2R$ to infinity. In the second domain $1/r$ is
used as a coordinate. For the $\phi$ dependence which is needed only
for the operator in~(\ref{eq:V}), 4 harmonics in $\phi$ are
sufficient.

\begin{figure}[htb]
    \centering \epsfig{file=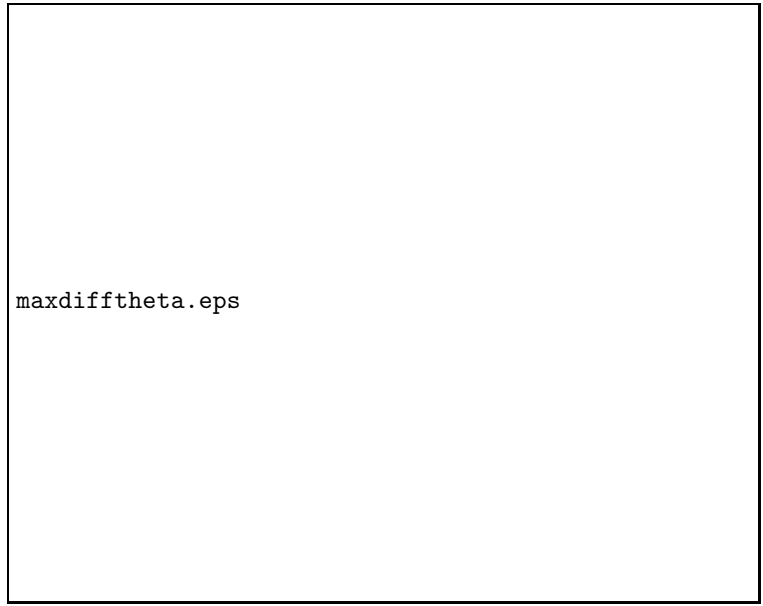,width=10cm}
    \caption{Difference between the exact and the numerical solution 
    for $\lambda=3$ and $\delta=0$ for $r=3$ in dependence on
    $\theta$.}
    \label{fig:maxdifftheta}
\end{figure}

\begin{figure}[htb]
    \centering \epsfig{file=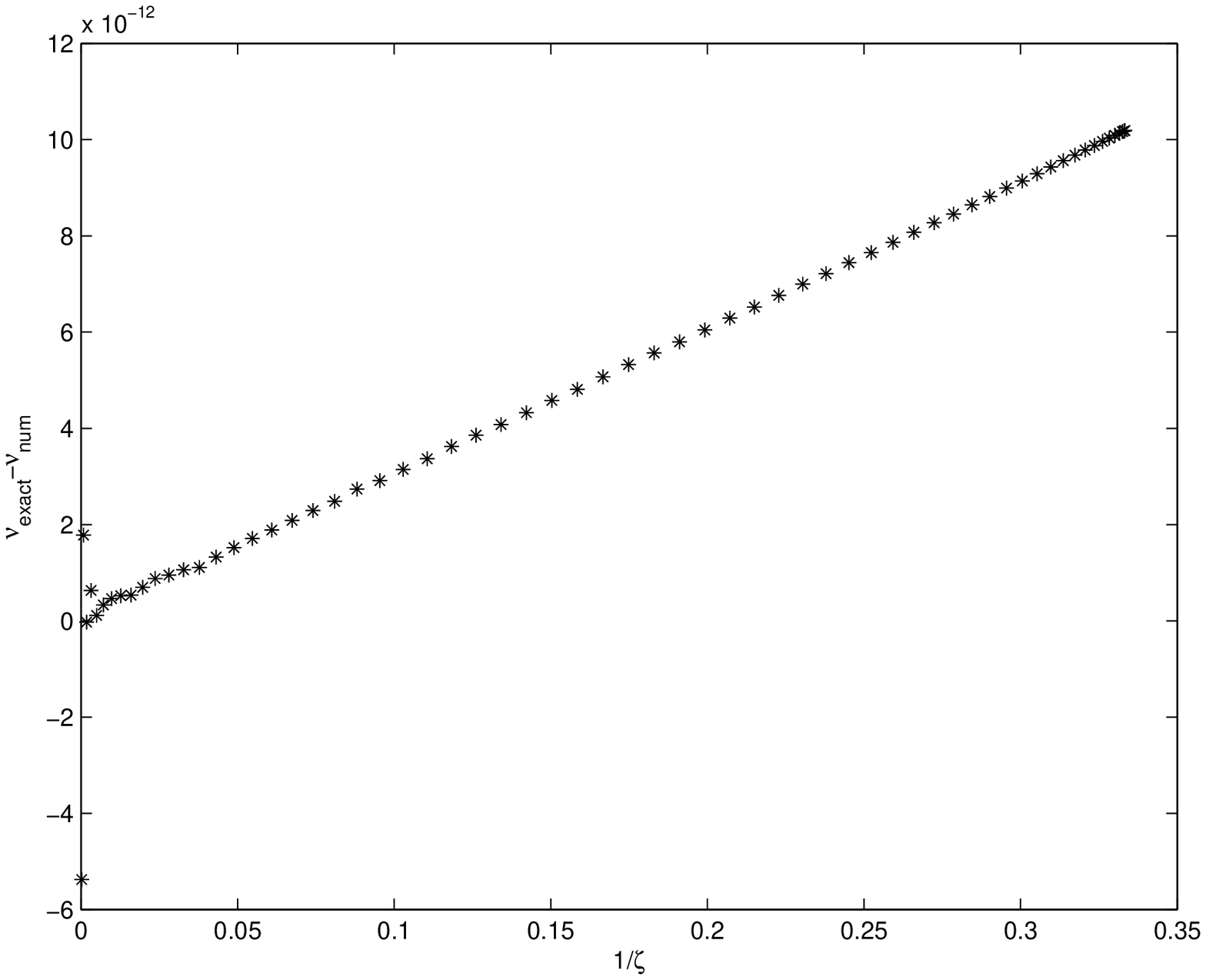,width=10cm}
    \caption{Difference between the exact and the numerical solution 
    for $\lambda=3$ and $\delta=0$ on the axis in dependence on $r$.}
    \label{fig:maxdiffr}
\end{figure}

Since \texttt{LORENE} is adapted to the solution of the Poisson equation, it is
to be expected that it reproduces the exact solution best for nearly
static situations, since the static solutions solve the Laplace
equation. The most significant deviations from the exact solution are
therefore expected for $\delta=0$. For the case $\lambda=3$, we
consider 32 harmonics in $\theta$ on a sphere of radius $R=1.5$. The
iteration is stopped if
$||\mathcal{F}_{n+1}-\mathcal{F}_{n}<5*10^{-10}$ which is the case in
this example after 90 steps. The exact solution is reproduced to the
order of $10^{-11}$. The absolute value of the difference
between the exact and the numerical solution on a sphere of radius 3
is plotted in Fig.~\ref{fig:maxdifftheta} in dependence of
$\theta$. There is no significant dependence of the error on $\theta$.
The maximal deviation is typically found on or near the axis.  As can
be seen from Fig.~\ref{fig:maxdiffr} which gives the dependence on $r$
on the axis, the error decreases almost linearly with $1/r$ except for
some small oscillations near infinity.

\begin{figure}[htb]
    \centering \epsfig{file=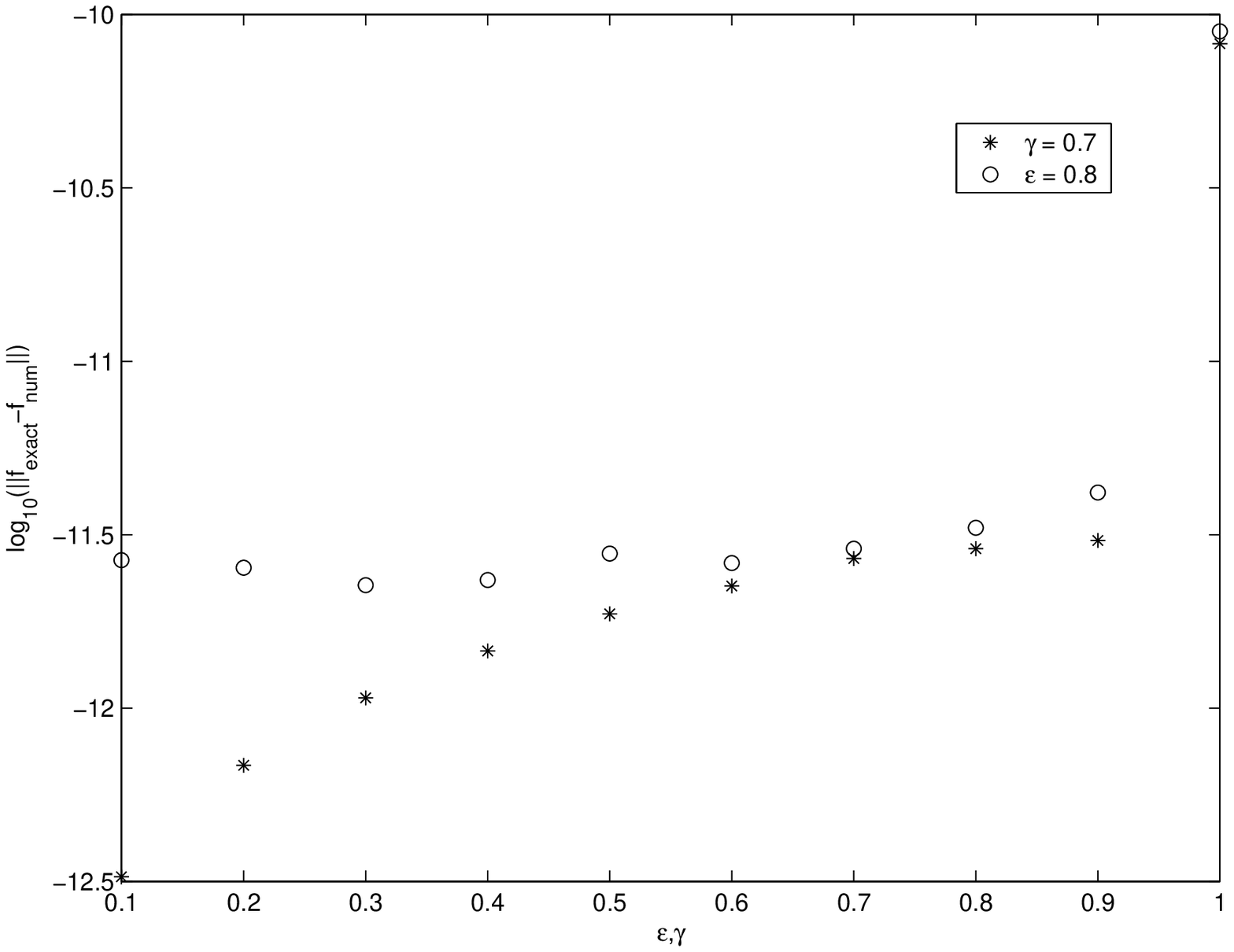,width=10cm}
    \caption{Difference between the exact and the numerical solution 
    for $\gamma=0.7$ and several values of $\epsilon$, and for
    $\epsilon=0.8$ and several values of $\gamma$.}
    \label{fig:gamma},
\end{figure}

We have plotted the maximal difference between the numerical and the
exact solution for a range of the physical parameters $\lambda$ and
$\delta$ in Fig.~\ref{fig:gamma}. As can be seen, the expectation is
met that the deviation from the exact solution increases if the
solution becomes more relativistic (larger $\epsilon$). As already
mentioned, the solution can be considered as exactly reproduced if the
deviation is below $10^{-13}$. Increasing the value of $\gamma$ for
fixed $\epsilon$ leads to less significant effects though the
solutions become less static with increasing $\gamma$.

For $\delta=0$, the ultra-relativistic limit $\lambda\to 4.629\ldots$
corresponds to a space-time with a singular axis which is not
asymptotically flat, see~\cite{prd4}. Since \texttt{LORENE} expands all
functions in a Galerkin basis with regular axis in an asymptotically
flat setting, solutions close to this singular limit cannot be
approximated. Convergence gets much slower and can only be achieved
with considerable relaxation. For $\lambda=4$ and $\delta=0$ we needed
nearly 2000 iterations with a relaxation parameter of
$\kappa=0.9$. The approximation is rather crude (in the order of one
percent). For higher values of $\lambda$ no convergence could be
obtained.

This is however due to the singular behavior of the solution in the
ultra-relativistic limit. In all other cases, \texttt{LORENE} is able to
reproduce the solution to the order of $10^{-11}$ and better, more
static and less relativistic cases are reproduced with the provided
accuracy.

\section{Conclusion}
\label{sec:concl}
In this article we have presented a scheme based on spectral methods
to treat hyperelliptic theta-functions numerically. It was shown that
an accuracy of the order of machine precision could be obtained with
an efficient code. As shown, spectral methods are very convenient if
analytic functions are approximated. Close to singularities such as
the degeneration of the Riemann surface, analytic techniques must be
used to end up with analytic integrands in the discussed example.

The obtained numerical data were used to provide boundary values for
the code \texttt{LORENE} which made possible a comparison of the
numerical solution to the boundary value problem with the numerically
evaluated theta-functions. For a large range of the physical
parameters the numerical solution was of the same quality as the
provided data. The main errors in \texttt{LORENE} are introduced by
rounding errors in the iteration. This shows that spectral methods
provide a reliable and efficient numerical treatment both for elliptic
equations and for hyperelliptic Riemann surfaces. However, to maintain
the global quality of the numerical approximation an analytical
understanding of the solutions is necessary in order to treat the
non-analyticities of the solutions.

\appendix
\section{Einstein equations and integral identities}
The Ernst equation has a geometric interpretation in terms of the
stationary axisymmetric Einstein equations in vacuum. The metric can
be written in this case in the Weyl-Lewis-Papapetrou form (see~\cite{exac})
\begin{equation}
    ds^{2}=g_{ab}dx^{a}dx^{b}=-f(dt+ad\phi)^{2}+(e^{2k}(d\rho^{2}+d\zeta^{2})
    +\rho^{2}d\phi^{2})/f
    \label{eq:wlp},
\end{equation}
where $\rho$ and $\zeta$ are Weyl's canonical coordinates and
$\partial_{t}$ and $\partial_{\phi}$ are the commuting asymptotically
timelike respectively spacelike Killing vectors.

In this case the vacuum field equations are equivalent to the Ernst
equation~(\ref{ernst1}) for the complex potential $\mathcal{E}$. For a
given Ernst potential, the metric~(\ref{eq:wlp}) can be constructed as
follows: the metric function $f$ is equal to the real part of the
Ernst potential. The functions $a$ and $k$ can be obtained via a line
integration from the equations
\begin{equation}
   a_{\xi}=2\rho\frac{(\mathcal{E}-\bar{\mathcal{E}})_{\xi}}{
   (\mathcal{E}+\bar{\mathcal{E}})^{2}}
    \label{axi},
\end{equation}
and
\begin{equation}
    k_{\xi}=(\xi-\bar{\xi})
    \frac{\mathcal{E}_{\xi}\bar{\mathcal{E}}_{\xi}}{
    (\mathcal{E}+\bar{\mathcal{E}})^{2}}\;.
    \label{kxi}
\end{equation}
This implies that $a$ is the dual of the imaginary part of the Ernst
potential. The equation~(\ref{kxi}) for $k$ follows from the equations
\begin{equation}
    R_{\alpha\beta}=\frac{1}{2f^{2}}\Re(\mathcal{E}_{\alpha}
    \bar{\mathcal{E}}_{\beta}),\quad \alpha,\beta=1,2,3
    \label{eq:ricci},
\end{equation}
where $R$ is the (three-dimensional) Ricci tensor corresponding to the
spatial metric $\mathrm{h}=\mbox{diag}(e^{2k},e^{2k},\rho^{2})$. This
reflects a general structure of the vacuum Einstein equations in the
presence of a Killing vector. For the Ricci scalar one finds
\begin{equation}
    -\frac{1}{2}e^{2k}R = k_{\rho\rho}+k_{\zeta\zeta}
    \label{virial17}.
\end{equation}
We denote by $h$ the determinant of the metric $\mathrm{h}$. 

The Komar integral \cite{komar,wald} of the twist of the timelike Killing vector 
$\xi=\partial_{t}$
over the whole spacetime establishes the equivalence between the 
asymptotically defined ADM mass and the Komar mass $m_{K}$, 
\begin{equation}
    2 
    \int_{disk}^{}dV\left(T_{ab}-\frac{1}{2}g_{ab}T^{c}_{c}\right)n^{a}
    \xi^{b}
    \label{virial2}=: m_{K},
\end{equation}
where the integration is carried out over the disk, 
 where $n_{a}$ is the normal at the 
disk, and where $T_{ab}$ is the energy momentum 
tensor of the disk given in \cite{prd4}.  In
other words the ADM mass can be calculated either asymptotically or
locally at the disk.

To obtain an identity which does not involve only surface integrals,
we consider as in~\cite{virial} an integral over the trace of
equation~(\ref{eq:ricci}) for the Ricci-tensor,
\begin{equation}
    R=\frac{h^{\alpha\beta}\mathcal{E}_{\alpha}
    \bar{\mathcal{E}}_{\beta}}{2f^{2}}
    \label{virial4}.
\end{equation}
To avoid numerical problems at the set of zeros of $f$, the so-called
ergo-sphere (see~\cite{prd4} for the disk solutions studied here), we
multiply both sides of equation~(\ref{eq:ricci}) by
$f^{3}$. Integrating the resulting relation over the whole space-time,
we find after partial integration
\begin{equation}
    -\int_{0}^{1}d\rho \rho f^{3}k_{\zeta}+
    \int_{0}^{\infty}d\rho\int_{-\infty}^{\infty}d\zeta ((\rho
    f^{3})_{\rho}k_{\rho}+(\rho f^{3})_{\zeta}k_{\zeta})=
    \int_{0}^{\infty}d\rho\int_{-\infty}^{\infty}d\zeta \rho f(
    \mathcal{E}_{\rho}\bar{\mathcal{E}}_{\rho}+
    \mathcal{E}_{\zeta}\bar{\mathcal{E}}_{\zeta})
    \label{eq:virial5};
\end{equation}
here the only contributions of a surface integral arise at the disk,
since $k\propto 1/r^{2}$ for $r\to\infty$ and since the axis is
regular ($k$ vanishes on the axis). If we replace $k$ via~(\ref{kxi}),
we end up with an identity for the Ernst potential and its
derivatives,
\begin{eqnarray}
   && -\int_{0}^{1}d\rho \rho^{2}f (\mathcal{E}_{\rho}
   \bar{\mathcal{E}}_{\zeta}
   +\mathcal{E}_{\zeta}\bar{\mathcal{E}}_{\rho})
   +\frac{3}{2}\int_{0}^{\infty}\int_{0}^{\infty}d\rho d\zeta
   \rho^{2}(\mathcal{E}_{\rho}(\bar{\mathcal{E}}_{\rho}^{2}
   +\bar{\mathcal{E}}_{\zeta}^{2})+\bar{\mathcal{E}}_{\rho}
   (\mathcal{E}_{\rho}^{2}+\mathcal{E}_{\zeta}^{2})) \nonumber\\ && =
   2\int_{0}^{\infty}\int_{0}^{\infty}d\rho d\zeta \rho f
   \mathcal{E}_{\zeta} \bar{\mathcal{E}}_{\zeta}
    \label{virial20}.
\end{eqnarray}
This identity (as the identity given in~\cite{virial}) can be seen as
a generalization of the Newtonian virial theorem. The
relation~(\ref{virial20}) coincides with the corresponding
relation of~\cite{virial} only in the Newtonian limit. This reflects
the fact that generalizations of a Newtonian result to a general relativistic
setting are not unique. Our formulation is adapted to the Ernst
picture and avoids problems at the ergo-spheres, thus it seems optimal
to test the numerics for Ernst potentials in terms of theta-functions.

\section*{Acknowledgment}
We thank A.~Bobenko, D.~Korotkin,
E.~Gourgoulhon and J.~Novak for helpful discussions and hints. CK is
grateful for financial support by the Marie-Curie program of the
European Union and the Schloessmann foundation.

\end{document}